\documentclass[journal]{IEEEtran}
\usepackage{amsmath,amsfonts}
\usepackage{algorithm}
\usepackage{algorithmic}
\usepackage{array}
\usepackage[caption=false,font=normalsize,labelfont=sf,textfont=sf]{subfig}
\usepackage{textcomp}
\usepackage{stfloats}
\usepackage{url}
\usepackage{verbatim}
\usepackage{graphicx}
\usepackage{cite}
\usepackage{diagbox}
\usepackage{amsmath}
\usepackage{multirow}
\usepackage{makecell}
\usepackage{setspace}
\usepackage{threeparttable}
\usepackage{color}
\usepackage{amsthm}

\usepackage{booktabs}
\usepackage{graphicx}

\def\BibTeX{{\rm B\kern-.05em{\sc i\kern-.025em b}\kern-.08em
    T\kern-.1667em\lower.7ex\hbox{E}\kern-.125emX}}
\usepackage{balance}

\begin{document}

\title{Admittance-Guided Inverter Dispatch Command Manipulation Attack:\\ A Grid Stability-Oriented Approach
}

\author{
Hongwei Zhen,~\IEEEmembership{Student Member,~IEEE,}
Ze Yu,~\IEEEmembership{Student Member,~IEEE,}
Xin Xiang,~\IEEEmembership{Member,~IEEE,} \\
Mingyang Sun,~\IEEEmembership{Senior Member,~IEEE,}
Wuhua Li,~\IEEEmembership{Senior Member,~IEEE}
\vspace{-30pt}

}



\maketitle

\begin{spacing}{0.9}

\begin{abstract}
The high penetration of voltage source converters in modern smart microgrids enhances operational flexibility while introducing complex cyber-physical vulnerabilities.  
Existing cyber-attack studies either require detailed knowledge of system topology and controller dynamics or depend on repeated online interactions, which may compromise practicality by generating operationally infeasible or limit-violating commands.
This article investigates a dispatch command manipulation attack and develops an admittance-guided framework to identify the vulnerable inverter and the worst-case dispatch command that most severely degrades system stability. A compromised inverter is utilized to inject controlled harmonic perturbations for sparse admittance measurement, and a physics-informed neural network is then employed to reconstruct the operating-point-dependent admittance of target inverters over the feasible dispatch region. Based on the reconstructed admittance, a stability-margin-oriented optimization is formulated to locate the most vulnerable inverter and the corresponding worst-case dispatch command. 
Controller hardware-in-the-loop experiments on a five-inverter microgrid demonstrate that the identified command can drive the system into severe sub-synchronous oscillations while remaining within nominal dispatch bounds, highlighting the need for stability-aware command screening beyond static limit checking.
\end{abstract}

\begin{IEEEkeywords}
smart microgrid cybersecurity, dispatch command manipulation attack, inverter-based resources, admittance measurement, physics-informed neural network.
\end{IEEEkeywords}

\vspace{-12pt}
\section{Introduction}

\IEEEPARstart{A}{s} modern microgrids become increasingly dominated by voltage source converters (VSCs), these inverters have become the primary interface for renewable generation and distributed energy resources. 
In contrast to synchronous-generator-based systems, VSC-dominated microgrids rely on fast digital control to regulate power exchange and maintain system-level performance. However, this technological transition also tightens the coupling between cyber access and physical dynamics. At the physical layer, the intrinsic low-inertia characteristics~\cite{background_Milano2018LowInertia} and nonlinear dynamics render the system highly sensitive to disturbances, narrowing stability margins~\cite{background_Wang2020GFMStability}. At the cyber layer, the programmability of inverters transforms them into high-risk targets, while enabling intelligent dispatch~\cite{review_sahoo2021}. As a result, inverters open a pathway for malicious adversaries to penetrate from cyberspace directly into the physical grid.

It is precisely this deep digitalization and network integration that has transformed modern microgrids into cyber-physical systems.
The threat landscape has expanded from centralized SCADA-targeted attacks, exemplified by the 2015 BlackEnergy3 incident~\cite{incident_Ukraine}, toward edge-side device exploitation. At the device level, Wi-Fi vulnerabilities discovered in Tigo Energy inverters~\cite{report_DERSecurity2024} enabled root-level access compromising residential power supply in 2016. At the system level, in 2017, the Horus attack~\cite{incident_Horus} scenario demonstrated that coordinated inverter manipulation could trigger cascading failures while the sPower denial-of-service attack~\cite{incident_sPower} caused a 12-hour communication blackout affecting 500 MW of renewable assets in 2019. Recent incidents reveal an evolution toward compound threats. Compromised monitoring platforms such as Solarman have enabled simultaneous user geolocation and firmware tampering~\cite{incident_SolarmanAPI}. In 2024, SolarView inverters manufactured by Contec were hijacked for financial fraud, marking the first confirmed cross-domain exploitation of inverter vulnerabilities~\cite{report_DERSecurity2024}.

These incidents demonstrate a clear threat trend, wherein attack targets have descended from upper-level control systems to underlying power electronic devices. Moreover, attack consequences have expanded from power interruptions to multidimensional threats encompassing system stability, user privacy, and property security. 
Confronted with this escalating security landscape, researchers have conducted systematic investigations spanning theoretical frameworks to empirical assessments. Existing works have established comprehensive threat models and resilience frameworks across device and grid-connected system levels~\cite{review_Liu2024, review_Yan2023}, while vulnerability analysis has identified potential attack pathways through firmware, network communications, and inverter control loops~\cite{review_sahoo2021}. To validate these theoretical insights, recent research has shifted toward experimental evaluation of commercial equipment using hardware-in-the-loop platforms~\cite{review_Musleh2024Experiment}, assessing vulnerability impacts on grid stability and economics. Nevertheless, critical examination of existing attack strategies reveals fundamental limitations in two key dimensions, the degree of system knowledge dependency and attack vector design.



Existing cyber-attack studies on inverter-dominated systems can be broadly categorized into privileged-access, physical-access, and model-based control-oriented approaches. At the cyber layer, denial-of-service and replay attacks~\cite{review_Yan2023}, control parameter tampering~\cite{traditional_Bamigbade2024PLL}, and firmware exploitation~\cite{traditional_Abraham2021Firmware} typically assume that adversaries have already obtained elevated system privileges. At the physical layer, hardware-targeted approaches such as Hall sensor spoofing~\cite{traditional_Barua2020Hall} and conducted electromagnetic interference on phase-locked loops~\cite{traditional_Albunashee2022PLL} require physical proximity to target equipment, imposing impractical deployment constraints. To exploit inverter closed-loop dynamics, more sophisticated paradigms have emerged. False data injection attacks (FDIA) have been extended to corrupt distribution network state estimation~\cite{whitebox_Zhuang2019SE} and mask anomalous photovoltaic behavior~\cite{whitebox_Liu2025Photovoltaic}. Eigenvalue-based signal design can excite latent electromechanical oscillation modes through wide-area damping control loops~\cite{whitebox_Wang2024MROCA}, while attacks on distributed secondary control disrupt power sharing~\cite{whitebox_Zhang2019LoadShare} and synchronization~\cite{whitebox_Mohamed2021SSOFDIA, whitebox_Jena2025PinningBased} in inverter-dominated microgrids. However, these model-based strategies universally presuppose complete knowledge of system topology, line impedance, and controller dynamics which constitutes an insurmountable barrier in practical scenarios.

To reduce dependence on detailed system knowledge, recent studies have increasingly explored interaction-driven and data-driven destabilization strategies. These methods typically formulate attack synthesis as a sequential decision-making problem, allowing learning agents to interact with the grid environment and search for destabilizing actions through trial and error. Representative examples include reinforcement-learning-based attacks on automatic generation control, load frequency control, and microgrid droop regulation~\cite{blackbox_Shereen2024AGC,blackbox_Said2025LFC,blackbox_Wang2023TD3}, as well as multi-agent false data injection strategies in distributed microgrids~\cite{blackbox_Abianeh2022Multiagent}. Related data-driven efforts have also examined dynamic load-altering attacks and fuzzing-assisted exploration of sensitive controller settings~\cite{blackbox_Amini2018DLAA,blackbox_Yu2025DARAFL}. 

Despite eliminating explicit model requirements, these approaches still exhibit two fundamental limitations that constrain their practical applicability. 
First, the learning-based attack process requires extensive online interaction with the physical grid. Agents must execute numerous exploratory episodes to achieve policy convergence, during which each probing action perturbs system states and may trigger protection relays, fault recorders, or power quality monitoring alarms~\cite{blackbox_Kurt2019Detection}. This tight coupling between learning and execution fundamentally makes offline attack planning infeasible, forcing adversaries to expose detectable signatures during the reconnaissance phase. 
Second, the absence of embedded power electronics knowledge leads to inefficient and often infeasible attack strategies, while stealthiness cannot be guaranteed.
Without awareness of inverter small-signal characteristics, learned policies cannot distinguish stability boundaries from normal operating regions. Consequently, generated commands may violate device thermal limits, grid code requirements, or protection coordination constraints~\cite{review_Liu2024}. Moreover, the conservative adjustment range of control parameters imposed by manufacturer firmware and protection settings restricts tampering-based attacks to narrow operating margins, fundamentally limiting their destabilization capability. Furthermore, the lack of stability-oriented guidance prevents systematic identification of the most vulnerable unit in multi-inverter systems, rendering indiscriminate attacks on arbitrary inverters ineffective even when full control access is available.

In summary, existing destabilization studies either rely on detailed knowledge of system topology and controller dynamics or depend on repeated online exploration without explicit stability-oriented physical guidance. This leaves a practical gap for methods that can characterize operating-point-dependent stability vulnerability with limited prior information. In VSC-dominated microgrids, the output admittance of a VSC varies with its steady-state operating point, which creates an opportunity for dispatch-command-induced migration toward low-damping regions. Motivated by this observation, this paper investigates a dispatch command manipulation attack and develops an admittance-guided identification framework under a limited-prior-knowledge setting. The proposed framework reconstructs operating-point-dependent admittance characteristics from sparse terminal measurements, identifies the most vulnerable inverter, and determines the worst-case dispatch command that minimizes the system stability margin.
The main contributions of this paper are summarized as follows:

\begin{itemize}

    \item To the best of our knowledge, this is the first work that incorporates admittance characteristics into the design of cyberattacks for VSC-dominated microgrids. An admittance-guided identification framework is developed for dispatch command manipulation attacks, revealing how dispatch-induced operating-point migration can drive inverters toward inherent low-damping regions and thereby degrade system stability margins, expanding the attack surface while reducing reliance on attack vectors that are readily exposed by conventional defenses based on static limit checking.
    
    \item A measurement-based admittance reconstruction strategy is established using limited network-level access. By repurposing one compromised inverter as a harmonic excitation source, sparse admittance samples are acquired from terminal measurements, and a physics-informed neural network (PINN) with transfer learning is employed to extrapolate these samples over the feasible operating region for stability-oriented analysis.
      
    \item A small-signal-stability-margin-oriented identification procedure is formulated to determine the most vulnerable inverter and the corresponding worst-case dispatch command. Controller hardware-in-the-loop (CHIL) experiments on a five-inverter microgrid further validate the effectiveness of the proposed framework in locating the most vulnerable inverter and determining the corresponding command that induces severe sub-synchronous oscillations (SSO).
\end{itemize}

\vspace{-10pt}
\section{Motivation and Threat Model}

\subsection{Motivation}

The distributed deployment of VSCs has led to widespread use of online remote control centers, substantially enlarging the cyber attack surface. Data from cyberspace search engines such as Shodan, Fofa, and Zoomeye reveal over 100,000 publicly accessible control nodes~\cite{blackbox_Yu2025DARAFL}. These centers typically feature web-based interfaces for remote command execution and parameter tuning. While enhancing interoperability, this design compromises security boundaries. The prevalent lack of robust access control and encryption renders these interfaces susceptible to Man-in-the-Middle attacks, exposing the control plane to persistent threats.

At the physical layer, inherent dynamic stability vulnerabilities arise from mismatches between universal control designs and specific engineering scenarios, introducing negative-damping at particular Steady-State Operating Points (SSOPs)~\cite{background_Wang2020GFMStability}. Moreover, modeling uncertainties from non-ideal component parameters and interaction effects in multi-inverter systems can intensify resonance or amplify instability modes. Consequently, even rigorously designed microgrids retain latent stability defects. An adversary may trigger instability by maliciously manipulating microgrid dispatch commands to steer inverter operating points toward inherent negative-damping regions within normal dispatch boundary, rather than relying on direct control logic or parameter tampering.

Existing studies have paid limited attention to this cross-layer threat mechanism, in which cyber access is used to activate inherent physical vulnerabilities, constituting the core motivation for this research. To address this issue, we develop an admittance-guided identification framework to characterize stability vulnerability and identify worst-case dispatch commands under adversarial manipulation.

\vspace{-1.5em}
\subsection{Threat Model} 

\begin{figure}[htbp]
    \centering
    \includegraphics[width=0.9\linewidth]{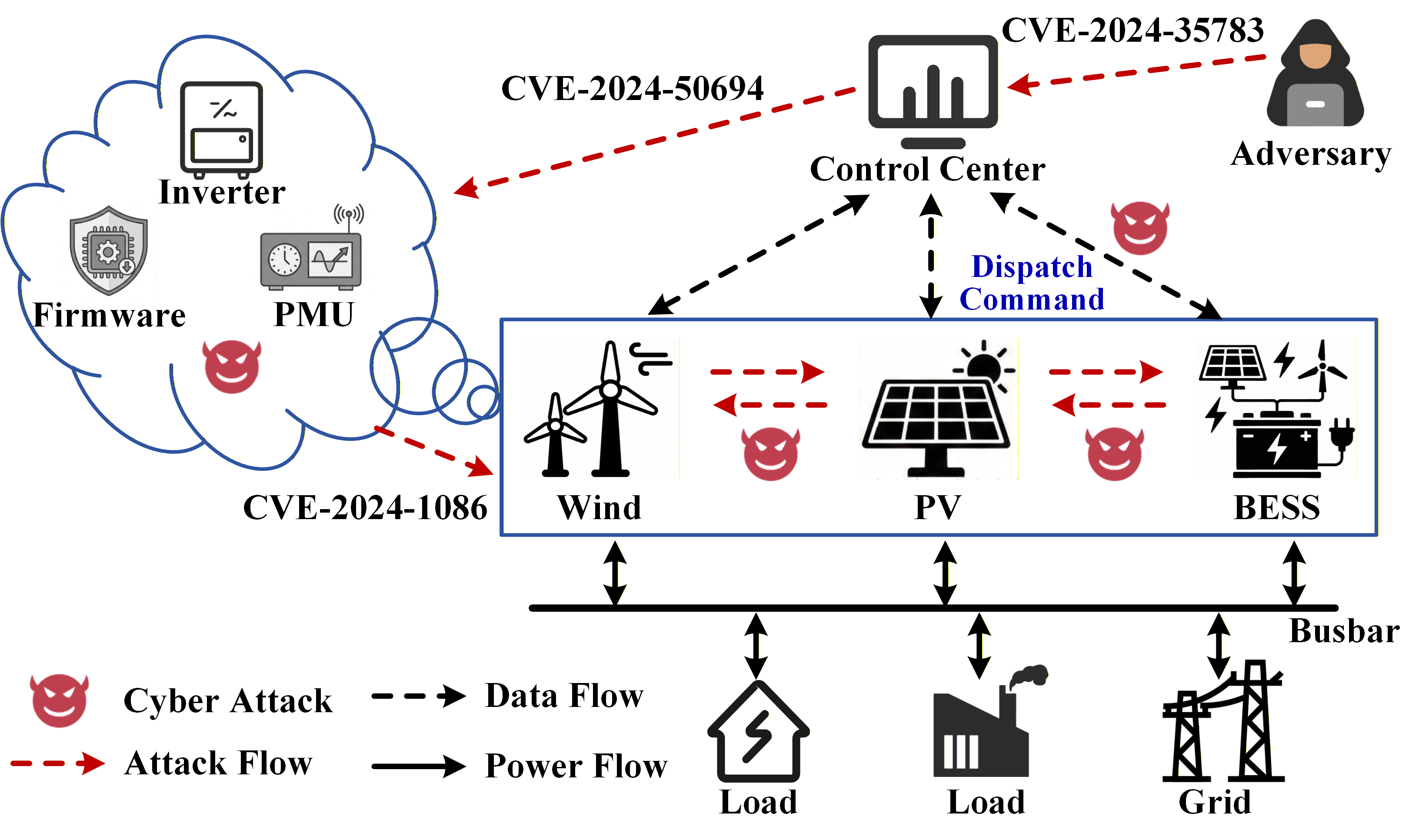}
    \vspace{-15pt}
    \caption{Cross-layer intrusion path from network-access vulnerabilities to physical-device manipulation.}
    \label{fig:threat model}
\end{figure}

In this paper, the cross-layer penetration path is illustrated in Fig.~\ref{fig:threat model}. 
Recent disclosures within the ecosystem of VSC-dominated microgrids substantiate the practical feasibility of the multi-stage infiltration, ranging from cloud-level access control bypass~\cite{cve_2024_35783} and gateway command injection~\cite{cve_2024_50694} to firmware privilege escalation~\cite{cve_2024_1086}.
Based on this landscape, the assumptions of adversary capabilities are given as follows:

\begin{enumerate}
\item \textit{Network Infiltration and Control}: The adversary has compromised the microgrid communication network and can modify dispatch commands while accessing partial internal controller registers of distributed energy resources. Consistent with commercial implementations, only commonly exposed parameters, such as proportional-integral gains and power setpoints, are assumed to be modifiable, whereas the exact topology and other core controller parameters remain unavailable~\cite{NIST8498}.
\item \textit{Active Harmonic Injection}: The adversary controls a single inverter to inject harmonic perturbations. Meanwhile, the adversary leverages the built-in oscillography and transient measurement functions of the target inverters to buffer high-resolution response waveforms synchronously, as required for electronically coupled resources~\cite{IEEE1547Std}. These locally stored datasets are subsequently exfiltrated via asynchronous communication to perform offline Fast Fourier Transform analysis.
\item \textit{Knowledge Constraints}: The adversary operates under a limited-prior-knowledge setting. Specifically, the adversary has no prior access to the exact system topology or full controller parameter set, and therefore relies on terminal measurements and the identified admittance characteristics for subsequent analysis.
\end{enumerate}

\vspace{-1em}

\vspace{-8pt}
These capabilities are intended to represent a security-evaluation upper bound combining commercially exposed control interfaces and experimentally realizable measurement access, rather than a universal assumption for all deployed microgrids.

\vspace{-9pt}
\subsection{Stability-Oriented Attack Mechanism}
Prior to elaborating on the proposed identification procedure, we first analyze the instability mechanism of grid-connected VSCs interacting with the grid based on admittance analysis. Under the small-signal framework, a VSC can be represented by its Norton equivalent circuit as a controlled source in parallel with an output admittance $\boldsymbol{Y}_{\mathrm{VSC}}(s, \boldsymbol{x}_{\mathrm{op}}, \boldsymbol{p})$, which is jointly determined by the controller parameters $\boldsymbol{p}$ and the steady-state operating points $\boldsymbol{x}_{\mathrm{op}}$. When the VSC is coupled with the grid, the system stability is governed by the minor-loop gain $\boldsymbol{L}$ formed by the VSC output admittance and the grid impedance $\boldsymbol{Z}_{\mathrm{g}}(s)$~\cite{Sun2011GNC}:
\begin{equation}
    \label{eq:General_GNC}
    \boldsymbol{L}(s, \boldsymbol{x}_{\mathrm{op}}, \boldsymbol{p}) = \boldsymbol{Z}_{\mathrm{g}}(s)\boldsymbol{Y}_{\mathrm{VSC}}(s, \boldsymbol{x}_{\mathrm{op}}, \boldsymbol{p})
\end{equation}

The generalized Nyquist criterion (GNC) is then applied to the return difference matrix
$\boldsymbol{I}+\boldsymbol{L}$,
where $\boldsymbol{I}$ denotes the identity matrix with compatible dimensions. If the interactive impedance exhibits negative damping characteristics within certain frequency bands, a positive feedback mechanism will be established, ultimately leading to oscillatory instability.

\begin{figure*}[htbp]
    \centering
    \includegraphics[width=0.9\linewidth]{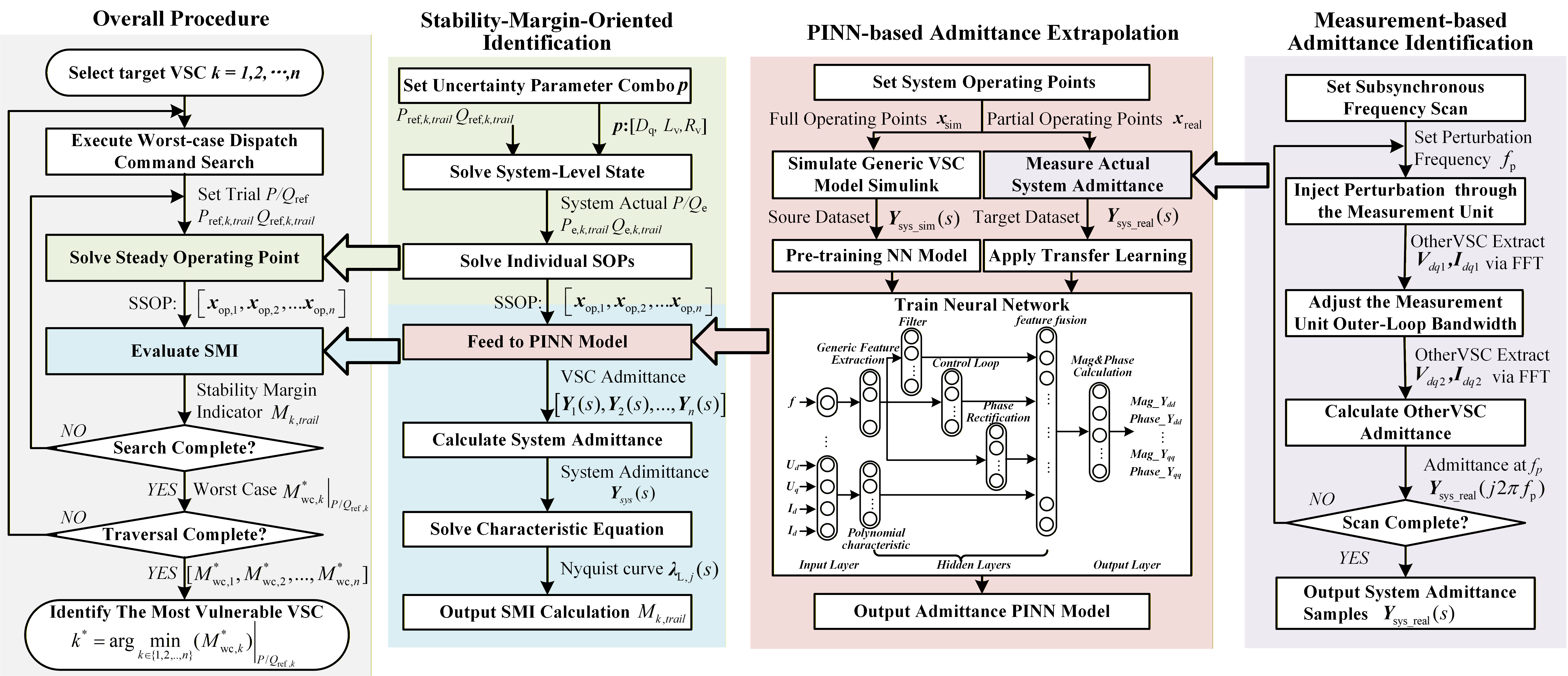}
    \vspace{-1em}
    \caption{The flowchart of the proposed admittance-guided identification procedure.}
    \label{fig:Flowchart}
    \vspace{-2em}
\end{figure*}

Existing destabilization attacks targeting VSC-dominated systems primarily inject disturbance signals or tamper controller parameters to artificially introduce negative damping, thereby undermining small-signal stability. However, these approaches overlook a critical fact that VSCs inherently possess low-damping vulnerable regions under varying operating points. Even when the controller parameters $\boldsymbol{p}$ remain unchanged, minor variations in the operating point $\boldsymbol{x}_{\mathrm{op}}$ can induce significant deviations in the admittance characteristics:
\begin{equation} 
    \label{eq:Admittance_OP}
    \Delta \boldsymbol{Y}_{\mathrm{VSC}}(s) = \frac{\partial{\boldsymbol{Y}_{\mathrm{VSC}}(s, \boldsymbol{x}_{\mathrm{op}})}}{\partial{\boldsymbol{x}_{\mathrm{op}}}} \cdot \Delta\boldsymbol{x}_{\mathrm{op}}
\end{equation}

The above analysis reveals a potential attack surface that, in certain off-nominal operating regions, VSCs may exhibit inherent negative-damping blind spots. 
To exploit these vulnerabilities, the adversary's objective is to pinpoint the destabilizing dispatch commands that force the system into the region with the lowest stability margin. Let $\boldsymbol{u} = [P_{\mathrm{ref}}, Q_{\mathrm{ref}}]^T$ denote the adversarial dispatch command vector and $\mathcal{O}$ represent the feasible command space. The attack strategy is formulated as the following constrained optimization problem:

\begin{equation}
\boldsymbol{u}^* = \mathop{\arg \min}_{\boldsymbol{u} \in \mathcal{O}} \quad \mathcal{M}(\boldsymbol{x}_{\mathrm{op}})
\quad\text{s.t.}\quad \boldsymbol{x}_{\mathrm{op}} = \mathcal{F}_{sys}(\boldsymbol{u})
\label{eq:attack_model}
\end{equation}

\noindent where $\mathcal{M}(\cdot)$ is the stability margin index evaluated from the GNC analysis of the system minor-loop gain, and $\boldsymbol{x}_{\mathrm{op}}$ denotes the steady-state operating point. The constraint $\mathcal{F}_{sys}(\cdot)$ represents the steady-state mapping determined by the microgrid power-flow and closed-loop control equations.

\vspace{-7pt}
\section{Admittance-Guided Identification Procedure for Dispatch Command Manipulation}
\vspace{-2pt}
\subsection{Overview of the Proposed Framework} 

Under the limited-prior-knowledge setting described in Section II, the adversary cannot determine a priori which inverter exhibits the lowest stability margin or which admissible dispatch command is most destabilizing. The resulting technical task is therefore to reconstruct the stability-relevant, operating-point-dependent admittance characteristics of target VSCs from accessible terminal measurements. Although the internal controller parameters are unavailable, terminal admittance can still be inferred from voltage and current measurements, and the feasibility of identifying admittance models from limited measurement samples has been demonstrated~\cite{Zhang2022PINN,Li2025PINN}.


As illustrated in Fig.~\ref{fig:Flowchart}, the proposed framework proceeds through three cascaded phases: measurement-based admittance identification, PINN-based admittance extrapolation, and identification of the most vulnerable inverter. The specific implementation and mathematical formulation of each phase are detailed in the subsequent subsections.


\vspace{-1em}
\subsection{Measurement-based Admittance Identification}

\begin{figure}[htbp]
    \centering
    \vspace{-1em}
    \includegraphics[width=0.95\linewidth]{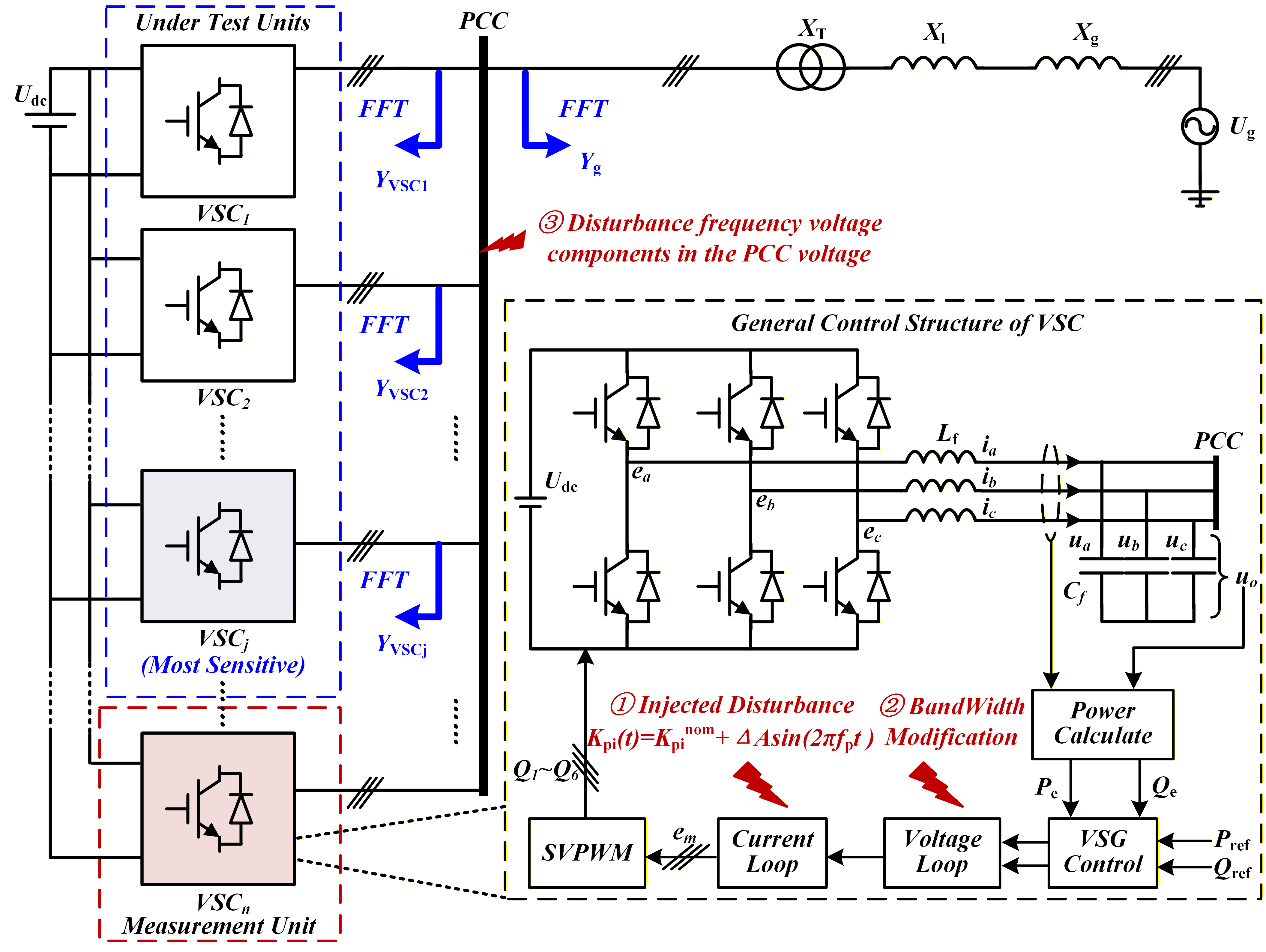}
    \vspace{-12pt}
    \caption{Using the Measurement-Unit-VSC for admittance measurements on the remaining UnderTest-Unit-VSCs within the microgrid.}
    \label{fig:SystemTopology}
    \vspace{-8pt}
\end{figure}

To acquire admittance data under the limited-prior-knowledge setting, a compromised VSC is repurposed as a Measurement Unit to actively probe the frequency response of the remaining Under-Test Units, as depicted in Fig.~\ref{fig:SystemTopology}. Conventional protection schemes primarily rely on static limit checking~\cite{Deng2017Defense}, with thresholds calibrated around normal operating conditions where stability margins are relatively high. Accordingly, direct parameter manipulation within these permissible bounds may not be sufficient to induce severe instability. Instead, the proposed procedure uses controlled parameter modulation to transform the Measurement Unit into a harmonic excitation source, thereby extracting terminal admittance data for subsequent vulnerability identification.


The small-signal dynamics of a Target Unit $k$ and the external grid interface in the Synchronous Reference Frame are governed by the admittance matrix $\boldsymbol{Y}_{k}(s)$ and the grid impedance $\boldsymbol{Z}_{g}(s)$, respectively. Since the perturbation injected by the Measurement Unit manifests as a common voltage excitation at the Point of Common Coupling (PCC), the identification of both parameters shares a unified linear regression formulation~\cite{Huang2009Measurement}. By defining the collected voltage response matrix as $\boldsymbol{U}_{\mathrm{mat}} = [\Delta \boldsymbol{U}_{dq}^{(1)}, \Delta \boldsymbol{U}_{dq}^{(2)}]$, the system equations at a specific frequency $f_p$ are expressed as:
\begin{equation}
    \begin{cases}
        [\Delta \boldsymbol{I}_{k}^{(1)}, \Delta \boldsymbol{I}_{k}^{(2)}] = \boldsymbol{Y}_{k}(j2\pi f_\mathrm{p}) \cdot \boldsymbol{U}_{\mathrm{mat}} \\
        \boldsymbol{U}_{\mathrm{mat}} = - \boldsymbol{Z}_\mathrm{g}(j2\pi f_\mathrm{p}) \cdot [\Delta \boldsymbol{I}_{g}^{(1)}, \Delta \boldsymbol{I}_{g}^{(2)}]
    \end{cases}
\end{equation}
where the superscripts $(1)$ and $(2)$ denote two distinct perturbation instances. The solution mandates that the voltage response matrix $\boldsymbol{U}_{\mathrm{mat}}$ satisfies the full-rank condition.
However, repetitive perturbation injections into a Linear Time-Invariant system operating at a fixed equilibrium inevitably yield linearly dependent response vectors, causing the rank deficiency of $\boldsymbol{U}_\mathrm{mat}$. To resolve this ill-posedness, the proposed procedure employs active impedance reshaping to ensure linear independence of the measurement vectors.

The proposed method uses a two-stage sequential perturbation mechanism. The excitation signal is synthesized internally by modulating the proportional gain $K_\mathrm{pi}$ of the inner current loop within the Measurement Unit:
\begin{equation}
\label{eq:Hamonic_Injection}
K_\mathrm{pi}(t) = K_\mathrm{pi}^\mathrm{nom} + \Delta A\sin(2\pi f_\mathrm{p}t)
\end{equation}
The perturbation amplitude $\Delta A$ is selected such that the modulated gain $K_\mathrm{pi}(t)$ remains within the permissible operational constraints of the controller. From a small-signal perspective, this modulation functions as an equivalent disturbance voltage source $\Delta \boldsymbol{u}_\mathrm{inj}$ injected at the current controller output.

The mechanism for generating linearly independent responses is derived from the closed-loop dynamics of the Measurement Unit. The transfer relationship governing the internal injection $\Delta \boldsymbol{u}_\mathrm{inj}$, terminal voltage $\Delta \boldsymbol{U}_{dq}$, and output current $\Delta \boldsymbol{I}_{dq}$ is expressed as:
\begin{equation} \label{eq:general_model}
\begin{split}
    \Delta \boldsymbol{u}_\mathrm{inj}(s) = & [\boldsymbol{Z}_\mathrm{f}(s) + \boldsymbol{G}_\mathrm{cc}(s)] \Delta \boldsymbol{I}_{dq} + \\
                          & [\boldsymbol{G}_\mathrm{cc}(s)\boldsymbol{G}_\mathrm{vc}(s) + \boldsymbol{I}] \Delta \boldsymbol{U}_{dq}
\end{split}
\end{equation}
where $\boldsymbol{Z}_\mathrm{f}$, $\boldsymbol{G}_\mathrm{cc}$, and $\boldsymbol{G}_\mathrm{vc}$ denote the filter impedance, inner current loop transfer matrix, and outer voltage loop transfer matrix, respectively. This reveals that the equivalent output admittance of the Measurement Unit is intrinsically dependent on the configuration of the outer loop controller $\boldsymbol{G}_\mathrm{vc}(s)$.

Leveraging this dependency, the proposed procedure ensures the non-singularity of $\boldsymbol{U}_\mathrm{mat}$ by exploiting the timescale separation inherent in VSC control architectures. During the first measurement stage, the Measurement Unit operates with a nominal voltage controller $\boldsymbol{G}_\mathrm{vc1}(s)$ and a bandwidth of $BW_1$. In the subsequent stage, the adversary modifies the outer loop parameters to transition to a secondary controller $\boldsymbol{G}_\mathrm{vc2}(s)$ configured with a distinct bandwidth $BW_2$. 

\begin{figure}[htbp]
    \centering
    \vspace{-1em}
    \includegraphics[width=0.9\linewidth]{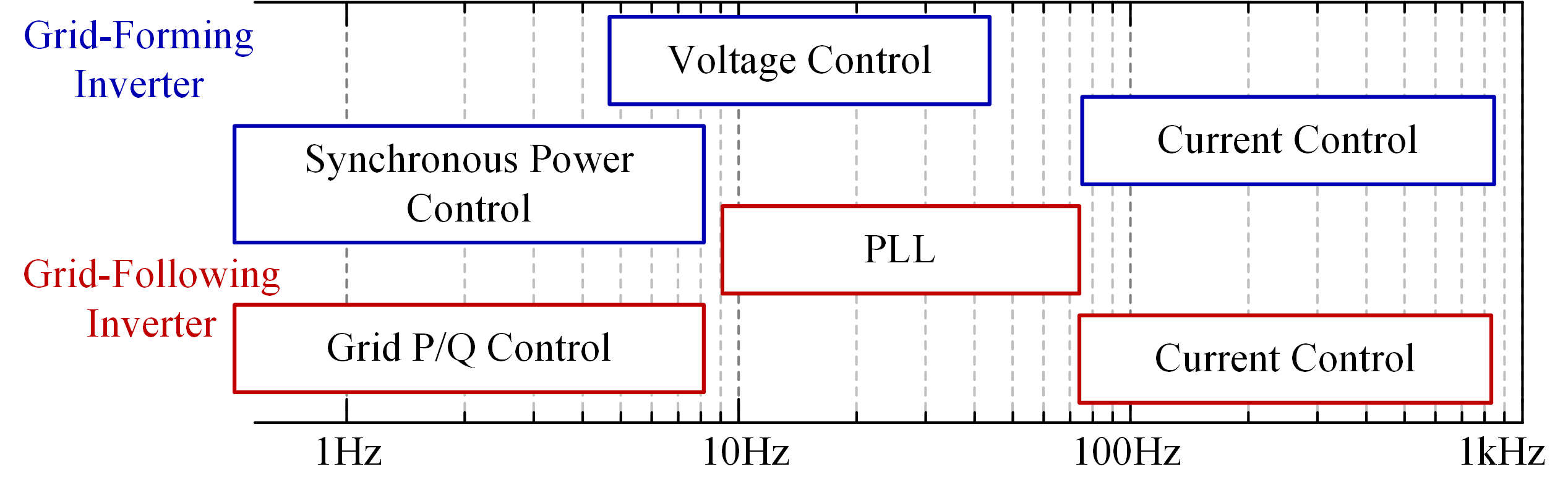}
    \vspace{-15pt}
    \caption{Comparison of different control loop bandwidth in grid-tied VSC control framework.}
    \label{fig:TimeScale}
    \vspace{-10pt}
\end{figure}

In practical engineering design, the bandwidth of the voltage control loop coincides with the sub-synchronous oscillations band as shown in Fig.~\ref{fig:TimeScale}. Consequently, the modification of $\boldsymbol{G}_\mathrm{vc}$ significantly reshapes the output admittance of the Measurement Unit at the perturbation frequency $f_p$. This impedance variation induces a rotation of the voltage vector in the dq-domain between the two stages, thereby ensuring the non-singularity. This process establishes a well-posed inverse problem and enables the unique identification of the admittance matrix $\boldsymbol{Y}_{dq,k}(j2\pi f_\mathrm{p})$.

\subsection{PINN-based Admittance Extrapolation}
\vspace{-10pt}
The measurement-based admittance identification described in Section~III.B yields discrete frequency-domain samples at limited stable operating points, whereas the subsequent vulnerability identification requires admittance evaluation over the entire operating region, including potentially unstable regions where direct measurement is infeasible. To bridge this gap, we develop a physics-informed neural network with transfer learning to extrapolate sparse measurements and reconstruct continuous admittance models over the full operating range. The proposed PINN architecture leverages a structured polynomial dependence of grid-tied VSC admittance on operating points, as established in prior work~\cite{Li2025PINN,Zhang2022PINN}. This structure is not introduced empirically, but follows directly from the linearization of the VSC system through the coupled action of three physical mechanisms.

The first source stems from the coordinate transformation. The synchronization loop provides the reference angle for the Park transformation. A small angular perturbation $\Delta\theta$ induces a geometric misalignment between the controller and system $dq$-frames:
\begin{equation}
\Delta \boldsymbol{i}_{dq}^{c} = \Delta \boldsymbol{i}_{dq}^{s} + \boldsymbol{i}_\theta \Delta\theta, \quad
\Delta \boldsymbol{u}_{dq}^{c} = \Delta \boldsymbol{u}_{dq}^{s} + \boldsymbol{u}_\theta \Delta\theta
\end{equation}
where superscripts $s$ and $c$ denote the system and controller frames, respectively. The coupling vectors
\begin{equation}
\boldsymbol{i}_\theta = \begin{bmatrix} I_q, -I_d \end{bmatrix}^T, \quad
\boldsymbol{u}_\theta = \begin{bmatrix} U_q, -U_d \end{bmatrix}^T
\end{equation}
depend linearly on the steady-state operating point $\boldsymbol{x}_{\mathrm{op}} = [U_d, U_q, I_d, I_q]^T$, thereby introducing steady-state variables into the small-signal equations.

The second source originates from the linearization of the power stage dynamics. In the average model of a VSC, the modulation indices couple the dc-link voltage to the ac-side filter states. During small-signal modeling around an equilibrium point, the steady-state modulation indices $\boldsymbol{M}_{dq0}$ appear as constant coefficients in the state-space matrices. Since $\boldsymbol{M}_{dq0}$ is rigorously determined by the terminal voltage and current $\boldsymbol{x}_{\mathrm{op}}$ to satisfy the steady-state voltage equations, the operating points are inherently embedded into the plant model parameters.

The third source is derived from the synchronization and power control loops. For control strategies such as Phase-Locked Loop(PLL), droop, and Virtual Synchronous Generator (VSG), the angular perturbation $\Delta\theta$ is calculated from terminal electrical quantities:
\begin{equation}
\Delta\theta = \boldsymbol{G}_{\mathrm{\theta i}}(s) \Delta \boldsymbol{i}_{dq} + \boldsymbol{G}_{\mathrm{\theta u}}(s) \Delta \boldsymbol{u}_{dq}
\end{equation}
Notably, the transfer function coefficients within $\boldsymbol{G}_{\mathrm{\theta i}}$ and $\boldsymbol{G}_{\mathrm{\theta u}}$ scale with the operating points. For instance, $\boldsymbol{G}_{\mathrm{\theta i}} \propto [U_d, U_q]^T$ and $\boldsymbol{G}_{\mathrm{\theta u}} \propto [I_d, I_q]^T$ in VSG control, with the proportionality determined by the power loop transfer function.

The multiplicative interaction among these three mechanisms generates the polynomial structure. When the coupling vectors from the coordinate transformation combine with the operating-point-dependent control coefficients, outer products emerge in the closed-loop transfer function. These interactions yield second-order terms such as $I_d U_q$ and $U_d^2$. Through systematic linearization of the complete VSC model, it has been shown~\cite{Liu2020Decouple} that the resulting admittance takes the form of a rational polynomial function:
\begin{equation}
    \label{eq:Y_poly}
    Y_{ml}(s, \boldsymbol{x}_{\mathrm{op}}) = \frac{\tilde{\boldsymbol{x}}^T \boldsymbol{P}_{ml}(s) \tilde{\boldsymbol{x}}}{\tilde{\boldsymbol{x}}^T \boldsymbol{P}_0(s) \tilde{\boldsymbol{x}}}, \quad m,l \in \{d,q\}
\end{equation}
where the augmented feature vector
\begin{equation}
    \label{eq:Xop}
    \tilde{\boldsymbol{x}} = [1, U_d, U_q, I_d, I_q, U_d^2, U_dU_q, U_dI_d, \ldots]^T
\end{equation}
contains polynomial terms of the operating point variables up to the second order and the coefficient matrices $\boldsymbol{P}_{ml}(s)$, $\boldsymbol{P}_0(s)$ encode the unknown system parameters including filter components and controller gains. This formulation achieves a critical decoupling: measurable operating points appear exclusively through $\tilde{\boldsymbol{x}}$, while unknown parameters are isolated in $\boldsymbol{P}(s)$.

\begin{figure}[htbp]
    \centering
    \vspace{-1.6em}
    \includegraphics[width=1\linewidth]{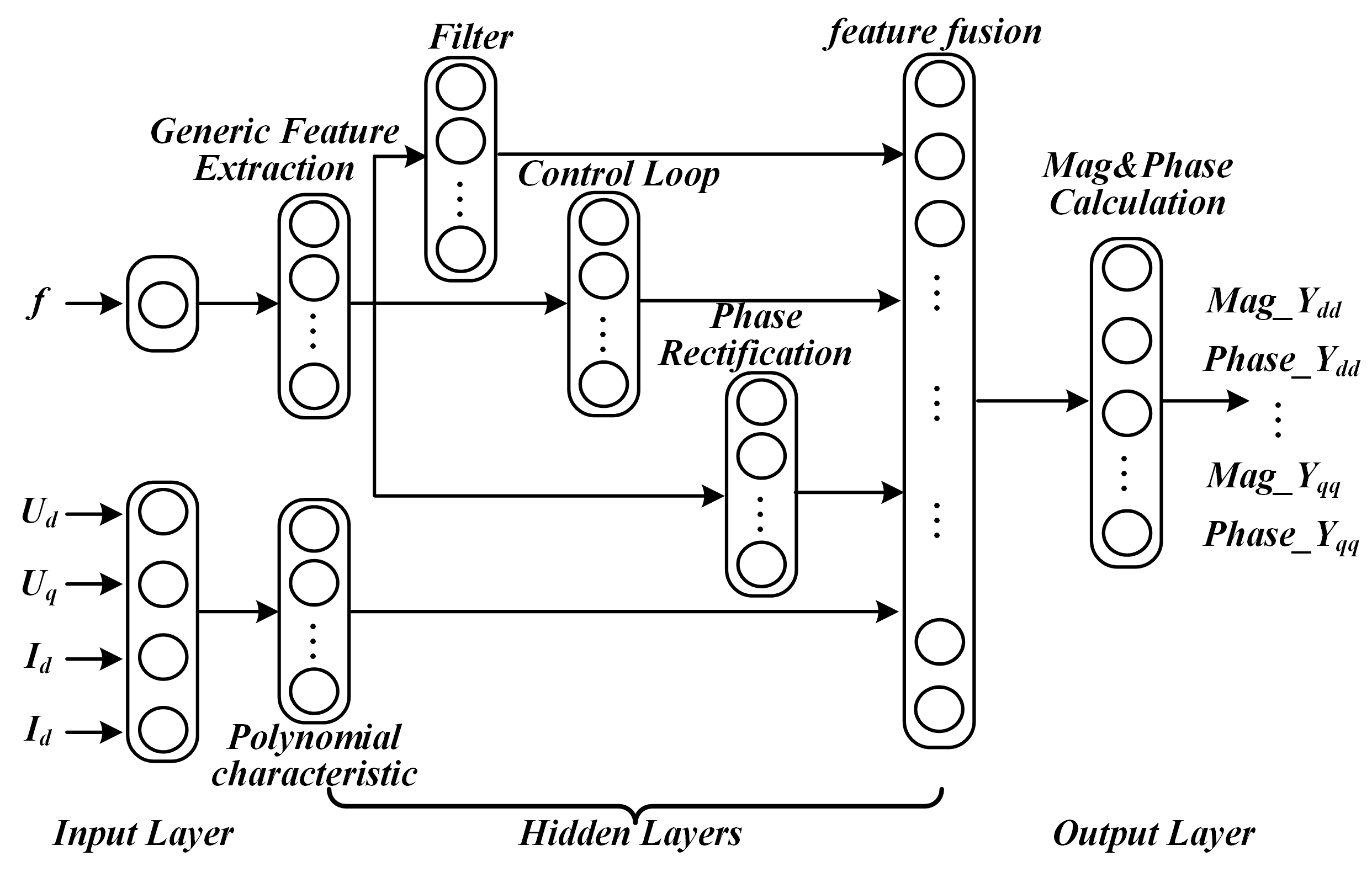}
    \vspace{-2em}
    \caption{Architecture of the proposed physics-informed neural network for admittance extrapolation.}
    \label{fig:NNStructure}
    \vspace{-1em}
\end{figure}

The derived polynomial formulation in \eqref{eq:Y_poly} serves as the structural blueprint for the PINN architecture shown in Fig.~\ref{fig:NNStructure}. To preserve physical consistency, the network adopts a dual-branch design with five functional layers. Layer~I lifts the operating-point variables into the augmented polynomial basis $\tilde{\boldsymbol{x}}$, while Layer~II learns a latent representation of the frequency-dependent coefficient matrices $\boldsymbol{P}(s)$. Layer~III further imposes physical modularity by disentangling the learned spectral features into subspaces associated with filter dynamics, control interactions, and coordinate transformation effects. Based on these representations, Layer~IV performs the quadratic fusion $\tilde{\boldsymbol{x}}^T \boldsymbol{P} \tilde{\boldsymbol{x}}$ to enforce physically valid admittance surfaces, and Layer~V maps the synthesized admittance elements into the complex plane to produce the magnitude and phase outputs required for the GNC stability assessment.

The network is trained by minimizing the mean squared error (MSE) between the predicted and measured admittance matrices:
\begin{equation}
\mathcal{L} = \frac{1}{N F_n} \sum_{i=1}^{N} \sum_{j=1}^{F_n} \|\boldsymbol{Y}_\mathrm{pred}(\boldsymbol{x}_{\mathrm{op},i}, f_j) - \boldsymbol{Y}_\mathrm{meas}(\boldsymbol{x}_{\mathrm{op},i}, f_j)\|_F^2
\label{eq:loss}
\end{equation}
where $N$ is the number of operating points and $F_n$ is the number of frequency samples. Dropout regularization after Layers~I and II enables adaptation to varying VSC complexities by pruning redundant features, with the optimal rate determined via Bayesian optimization.

To further improve sample efficiency, we adopt a transfer-learning strategy. The network is first pre-trained on a source dataset generated from analytical admittance models and is then fine-tuned using sparse measurements from target VSCs. Based on representational similarity analysis across layers, the transfer protocol freezes layers associated with universal feature extraction and updates only the control-loop-specific layers using the limited target dataset.

\vspace{-12pt}
\subsection{Identification of the Most Vulnerable Inverter}

With the PINN-based admittance model established in Section~III.C, the VSC admittance $\boldsymbol{Y}_{\mathrm{VSC}, i}(s, \boldsymbol{x}_{\mathrm{op}})$ can be evaluated at any operating point driven by feasible dispatch commands. Building upon this capability, this section develops a stability-margin-oriented identification procedure to determine the most vulnerable inverter and its corresponding worst-case dispatch command.

\begin{figure}[htbp]
    \centering
    \vspace{-15pt}
    \includegraphics[width=1\linewidth]{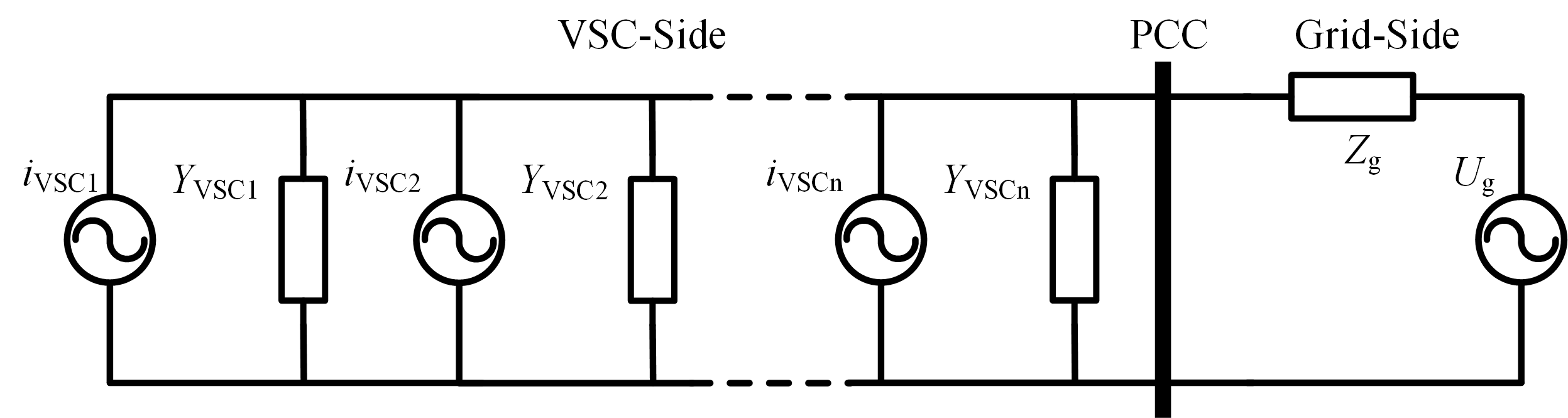}
    \vspace{-2em}
    \caption{Norton-equivalent representation of the grid-tied microgrid.}
    \label{fig:EqualizationCircuit}
    \vspace{-1em}
\end{figure}

Consider a microgrid comprising $n$ VSCs connected in parallel at PCC, which interfaces with the external grid through impedance $\boldsymbol{Z}_\mathrm{g}$. Applying Norton equivalence to the grid-tied system as illustrated in Fig.~\ref{fig:EqualizationCircuit}, the PCC voltage is governed by
\begin{equation}
    \boldsymbol{U}_\mathrm{pcc} = \left(\boldsymbol{I} + \boldsymbol{Z}_\mathrm{g} \sum_{i=1}^{n} \boldsymbol{Y}_{\mathrm{VSC}, i}\right)^{-1} \left(\boldsymbol{Z}_\mathrm{g} \sum_{i=1}^{n} \boldsymbol{I}_{VSC,i} + \boldsymbol{U}_\mathrm{g}\right)
\end{equation}
where $\boldsymbol{Y}_{\mathrm{VSC}, i} \in \mathbb{R}^{2 \times 2}$ denotes the complete admittance matrix of the $i$-th VSC obtained from the trained PINN model, and $Y_{ml}$ represents its elements. $\boldsymbol{I}_{\mathrm{VSC},i}$ represents its current injection, and $\boldsymbol{U}_\mathrm{g}$ is the grid voltage. The system stability is determined by the poles of the first term, motivating the definition of the minor-loop gain matrix
\begin{equation}
    \boldsymbol{L}_{sys}(s) = \boldsymbol{Z}_\mathrm{g}(s) \sum_{i=1}^{n} \boldsymbol{Y}_{\mathrm{VSC}, i}(s, \boldsymbol{x}_{\mathrm{op},i})
\end{equation}

Stability is analyzed using the Generalized Nyquist Criterion applied to the eigenvalues $\lambda_{\mathrm{L},j}(s)$ of $\boldsymbol{L}_{sys}(s)$. Under the limited-prior-knowledge setting, explicit admittance expressions are unavailable, and the trained PINN is therefore used to evaluate $\boldsymbol{Y}_{\mathrm{VSC}, i}(j\omega)$ at discrete frequencies within the sub-synchronous oscillations band $\Omega_{SSO}$. The corresponding eigenvalues are obtained by solving:
\begin{equation}
    \det(\boldsymbol{L}_{sys}(j2\pi f) - \lambda \boldsymbol{I}) = 0, \quad \forall f \in \Omega_{SSO}
\end{equation}

To quantify the proximity to instability, we define a Stability Margin Index (SMI) that combines directional information with distance measurement,assuming no right-half-plane poles in the open-loop transfer matrix. The sign component $S_\mathrm{k}$ indicates whether any eigenvalue trajectory encircles the critical point:
\begin{equation}
    S_\mathrm{k} = 
    \begin{cases}
    -1,  \text{if } \exists \lambda_{\mathrm{L},j}(j2\pi f) \text{ encircling } (-1, 0), \ \forall f \in \Omega_{SSO} \\
    +1,  \text{otherwise}
    \end{cases}
\end{equation}
The magnitude component $d_\mathrm{k}$ measures the minimum distance from any eigenvalue trajectory to the critical point:
\begin{equation}
    d_\mathrm{k} = \min_{j=1,2, f \in \Omega_{SSO}} \|\lambda_{\mathrm{L},j}(j2\pi f) + 1\|_2
\end{equation}

The composite SMI is defined as
\begin{equation}
    M_k = S_\mathrm{k} \cdot d_\mathrm{k}
\end{equation}
where negative values indicate instability and the magnitude quantifies the stability reserve or deficit. Smaller positive values correspond to reduced stability margins, making the system more susceptible to oscillations, while more negative values represent more severe instability as eigenvalues penetrate deeper beyond the critical point $(-1, 0)$.

Under the limited-prior-knowledge setting, the internal control parameters of each VSC remain unavailable. These parameters include the Q-V droop coefficients $D_{\mathrm{q},i}$ and virtual impedances $(R_{\mathrm{v},i}, L_{\mathrm{v},i})$, which govern the nonlinear mapping from dispatch commands $\boldsymbol{u}_k = [P_{\mathrm{ref},i}, Q_{\mathrm{ref},i}]^T$ to the actual steady-state operating point. Although precise values are unknown, their feasible ranges can be estimated based on typical design practices, forming the uncertainty set $\mathcal{P}$. The specific parameter bounds used in this work are detailed in Section~IV.

Based on this uncertainty description, we formulate a worst-case optimization to identify the dispatch commands that minimize the SMI across all plausible parameter realizations. For each VSC $k$, the optimization problem is expressed as
\begin{equation}
    \label{eq:optimization}
    M_{\mathrm{wc},k}^{*} = \min_{\boldsymbol{u}_k \in \mathcal{O}_k} \left\{ \min_{\boldsymbol{p} \in \mathcal{P}} M_k(\boldsymbol{u}_k, \boldsymbol{p}) \right\}
\end{equation}
where $\mathcal{O}_k$ denotes the feasible command space for VSC $k$, and $\boldsymbol{p} = [D_{\mathrm{q},1}, \ldots, D_{\mathrm{q},n}, R_{\mathrm{v},1}, \ldots, L_{\mathrm{v},n}]^T$ is the aggregated uncertain parameter vector.

Evaluating $M_k(\cdot)$ requires the actual steady-state operating point associated with a given dispatch command. This mapping is complicated by the droop characteristics, which cause the actual power output to deviate from the commanded values. For a given command pair $(P_{\mathrm{ref},k}, Q_{\mathrm{ref},k})$ and parameter instance $\boldsymbol{p}$, the Q-V droop imposes coupling between the PCC voltage magnitude $U_{\mathrm{pcc}}$ and total reactive power injection. The system-level power balance satisfies
\begin{equation}
\begin{cases}
Q_{\mathrm{pcc}} = \displaystyle\sum_{i=1}^{n} \left[ Q_{\mathrm{ref},i} + D_{\mathrm{q},i}(U_{\mathrm{nom}} - U_{\mathrm{pcc}}) \right] \\[8pt]

U_{\mathrm{pcc}} = U_{\mathrm{nom}} + (\displaystyle\sum_{i=1}^{n} P_{\mathrm{ref},i} \cdot R_\mathrm{g} + Q_{\mathrm{pcc}} \cdot X_\mathrm{g})/{U_{\mathrm{nom}}}

\end{cases}
\end{equation}
where $U_{nom}$ denotes the nominal voltage magnitude. Solving this coupled system yields $U_{\mathrm{pcc}}$ and the actual reactive power output of each VSC: $Q_{\mathrm{e},i} = Q_{\mathrm{ref},i} + D_{\mathrm{q},i}(U_{\mathrm{nom}} - U_{\mathrm{pcc}})$.

Once the voltage magnitude and reactive power are determined, the steady-state operating point of each VSC is characterized by the dq-frame variables $\boldsymbol{x}_{\mathrm{op},i} = \{U_{d,i}, U_{q,i}, I_{d,i}, I_{q,i}\}$. These are obtained by solving the power balance and virtual impedance equations:
\begin{equation}
\begin{cases}
    \label{eq:steady_state}
    P_{\mathrm{e},i} = U_{d,i} I_{d,i} + U_{q,i} I_{q,i} \\
    Q_{\mathrm{e},i} = U_{q,i} I_{d,i} - U_{d,i} I_{q,i} \\
    U_{\mathrm{pcc}}^2 = U_{d,i}^2 + U_{q,i}^2 \\
    U_{d,i} = E_{\mathrm{v}}-R_{\mathrm{v},i} I_{d,i} + \omega_n L_{\mathrm{v},i} I_{q,i} \\
    U_{q,i} = -R_{\mathrm{v},i} I_{q,i} - \omega_n L_{\mathrm{v},i} I_{d,i}
\end{cases}
\end{equation}
where $\omega_n$ is the nominal angular frequency, $P_{\mathrm{e},i} = P_{\mathrm{ref},i}$ under the assumption of ideal active power tracking and $E_{\mathrm{v}}$ denotes the common virtual electromotive-force magnitude of the VSCs. This nonlinear system is solved numerically using Newton-Raphson iteration for each parameter realization.

The identification procedure then proceeds through a nested evaluation over the feasible command space $\mathcal{O}_k$ and the uncertainty set $\mathcal{P}$. For each candidate command pair $(P_{\mathrm{ref},k}, Q_{\mathrm{ref},k})$ and parameter sample $\boldsymbol{p}$, the coupled droop and power-balance equations are first solved to obtain the steady-state operating points. These operating points are then fed into the trained PINN model to evaluate the admittance matrices across $\Omega_{SSO}$. Next, the loop-gain matrix $\boldsymbol{L}_{sys}(j2\pi f)$ is constructed, its eigenvalues are computed at each frequency, and the corresponding Nyquist loci are traced. The SMI is then evaluated from the proximity of the eigenvalue trajectories to the critical point. By sweeping the discretized set $\mathcal{O}_k \times \mathcal{P}$, the worst-case SMI $M_{\mathrm{wc},k}^{*}$ for each VSC is obtained as the minimum value encountered.

Finally, the most vulnerable inverter $k^{*}$ is identified as the unit yielding the globally minimum worst-case SMI:
\begin{equation}
    \label{eq:SensitiveVSC}
    k^{*} = \arg\min_{k \in \{1, \ldots, n\}} M_{\mathrm{wc},k}^{*}
\end{equation}
The corresponding commands $\boldsymbol{u}_k^{*}=[P_{\mathrm{ref},k^*}^{*}, Q_{\mathrm{ref},k^*}^{*}]^T$ constitute the worst-case dispatch commands that, when injected into VSC $k^{*}$, induce maximum degradation of the system stability margin while remaining within nominal operating bounds.

\vspace{-1em}
\section{Experiment Validation}  
\subsection{Experiment Setup}

Case studies are conducted to validate the proposed framework and to assess the impact of the identified dispatch commands on smart-microgrid stability. All VSCs in the CHIL platform are implemented as four-quadrant bidirectional converters. Therefore, the admissible dispatch-command search domain $(P_\mathrm{ref},Q_\mathrm{ref}) \in [-1,1]$ p.u. is physically supported by the experimental testbed.

\begin{figure}[h]
    \centering
    \vspace{-2em}
    \includegraphics[width=0.9\linewidth]{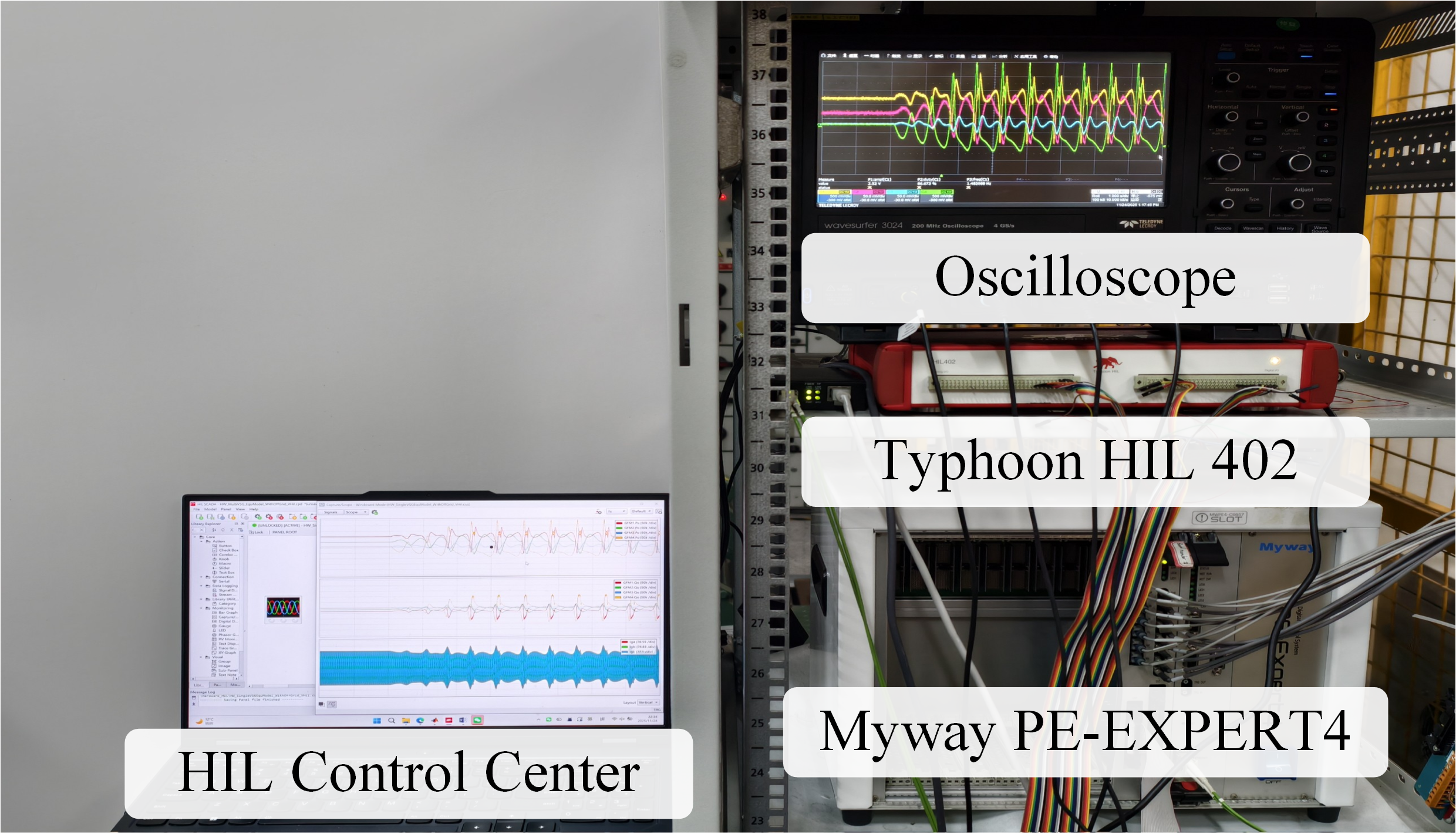}
    \vspace{-1em}
    \caption{Configuration of the CHIL experimental platform.}
    \label{fig:CHIL}
    \vspace{-2em}
\end{figure}

The proposed framework is validated through MATLAB/Simulink simulations and real-time controller hardware-in-the-loop experiments on the five-VSC microgrid depicted in Fig.~\ref{fig:SystemTopology}. The PINN algorithm is implemented in Python 3.9 and trained on an NVIDIA A100 GPU, requiring approximately 20 minutes to complete the full analysis. Table~\ref{tab:vsc_parameters} lists the parameters of the five VSCs in the test microgrid, which are designed following established VSG tuning guidelines in~\cite{Wu2016Paradesign}. The test system covers a range of power ratings and filter configurations representative of practical microgrids.
Real-time validation is conducted on a CHIL platform as shown in Fig.~\ref{fig:CHIL}, where the microgrid power stage is emulated on a Typhoon HIL 402 and the control algorithms are executed on Myway PE-EXPERT4 digital signal processors.

\begin{table}[h]
\centering
\vspace{-15pt}
\caption{Parameters of VSCs in microgrid}
\label{tab:vsc_parameters}
\vspace{-0.7em}
\scriptsize
\setlength{\tabcolsep}{1.4pt}
\renewcommand{\arraystretch}{1.12}
\begin{tabular}{@{}
    >{\centering\arraybackslash}p{1.35cm}
    >{\centering\arraybackslash}p{2.9cm}
    @{\hspace{0.7pt}}
    c
    @{\hspace{0.6pt}}
    c
    @{\hspace{0.6pt}}
    c
    @{\hspace{0.6pt}}
    c
    @{\hspace{0.6pt}}
    c
    @{}}
\hline\hline
& \textbf{Parameter} & \textbf{VSC-A} & \textbf{VSC-B} & \textbf{VSC-C} & \textbf{VSC-D} & \textbf{VSC-E} \\
\hline

\multirow{6}{*}{\makecell[c]{\textbf{Hardware}}}
& \makecell[c]{Rated Power $S_\mathrm{N}$ (kVA)} & 100 & 50 & 10 & 10 & 10 \\
& \makecell[c]{DC Voltage $U_{\mathrm{dc}}$ (V)} & \multicolumn{5}{c}{800} \\
& \makecell[c]{Grid Impedance} & \multicolumn{5}{c}{$L_g = 800\,\mu\mathrm{H},\; R_g = 0.1\,\Omega$} \\
& \makecell[c]{Filter Inductance\\$L_\mathrm{f}$ (mH)} & 0.8 & 1.6 & 4.8 & 3.2/0.8(LCL) & 4.8 \\
& \makecell[c]{Filter Capacitance\\$C_\mathrm{f}$ ($\mu$F)} & 200 & 100 & 50 & 50 & 50 \\
& \makecell[c]{Inertia Coefficient\\$J$ (kg$\cdot$m$^2$)} & 0.16 & 0.081 & 0.041 & 0.054 & 0.041 \\

\multirow{3}{*}{\makecell[c]{\textbf{Power}\\\textbf{Outer Loop}}}
& \makecell[c]{Active Power Damping\\$D_\mathrm{p}$ (kW/(rad/s))} & 200 & 120 & 32 & 16 & 32 \\
& \makecell[c]{Reactive Voltage-Droop\\Coefficient $D_\mathrm{q}$ (A)} & 3210 & 1605 & 321 & 321 & 321 \\
& \makecell[c]{Reactive Inertia\\Coefficient $K$ (A$\cdot$s)} & 71 & 35.5 & 7.1 & 7.1 & 7.1 \\

\multirow{2}{*}{\makecell[c]{\textbf{Voltage} \textbf{Loop}}}
& \makecell[c]{Proportional Gain $k_{\mathrm{pv}}$} & 1.5 & 1 & 0.4 & 1.2 & 0.4 \\
& \makecell[c]{Integral Gain $k_{\mathrm{iv}}$} & 900 & 450 & 90 & 90 & 90 \\

\multirow{1}{*}{\makecell[c]{\textbf{Current} \textbf{Loop}}}
& \makecell[c]{Proportional Gain $k_{\mathrm{pi}}$} & 3.5 & 7 & 9 & 11 & 9 \\

\multirow{2}{*}{\makecell[c]{\textbf{Virtual}\\\textbf{Impedance}}}
& Virtual Resistance $R_\mathrm{v}$ ($\Omega$) & 0 & 0.6 & 0.8 & 2.4 & 0.8 \\
& Virtual Inductance $L_\mathrm{v}$ (mH) & 0.4 & 0.4 & 1.6 & 1 & 1.6 \\

\hline\hline
\vspace{-5pt}
\end{tabular}
\end{table}

\vspace{-3em}
\subsection{Validation of the Proposed Identification Procedure}
This subsection presents the identification of the most vulnerable inverter and the corresponding worst-case dispatch command associated with the minimum stability margin.

\subsubsection{Performance of Measurement-Based Admittance Identification}

The measurement procedure follows the two-stage protocol described in Section III-B. In both stages, harmonic perturbations are injected by modulating the current-loop proportional gain of VSC-E according to \eqref{eq:Hamonic_Injection}. Between the two stages, the voltage-loop bandwidth of VSC-E is adjusted from its nominal value to a secondary configuration, thereby altering its equivalent output admittance and preserving the non-singularity of the voltage-response matrix.

Admittance is characterized using logarithmically spaced frequencies $f_\mathrm{p} \in [2, 130]$~Hz to capture sub-synchronous oscillations dynamics. To maintain operational plausibility, the measurements are restricted to a discretized dispatch-command grid defined by $P_{\mathrm{ref}} \in [0.2, 1.0]$~p.u.\ and $Q_{\mathrm{ref}} \in [-0.4, 0.4]$~p.u.\ with 0.2~p.u.\ increments. This configuration yields 25 distinct steady-state operating points, covering the normal operating envelope while excluding impractical low-power conditions.


\begin{figure}[hbtp]
    \centering
    \vspace{-15pt}
    \includegraphics[width=0.9\linewidth]{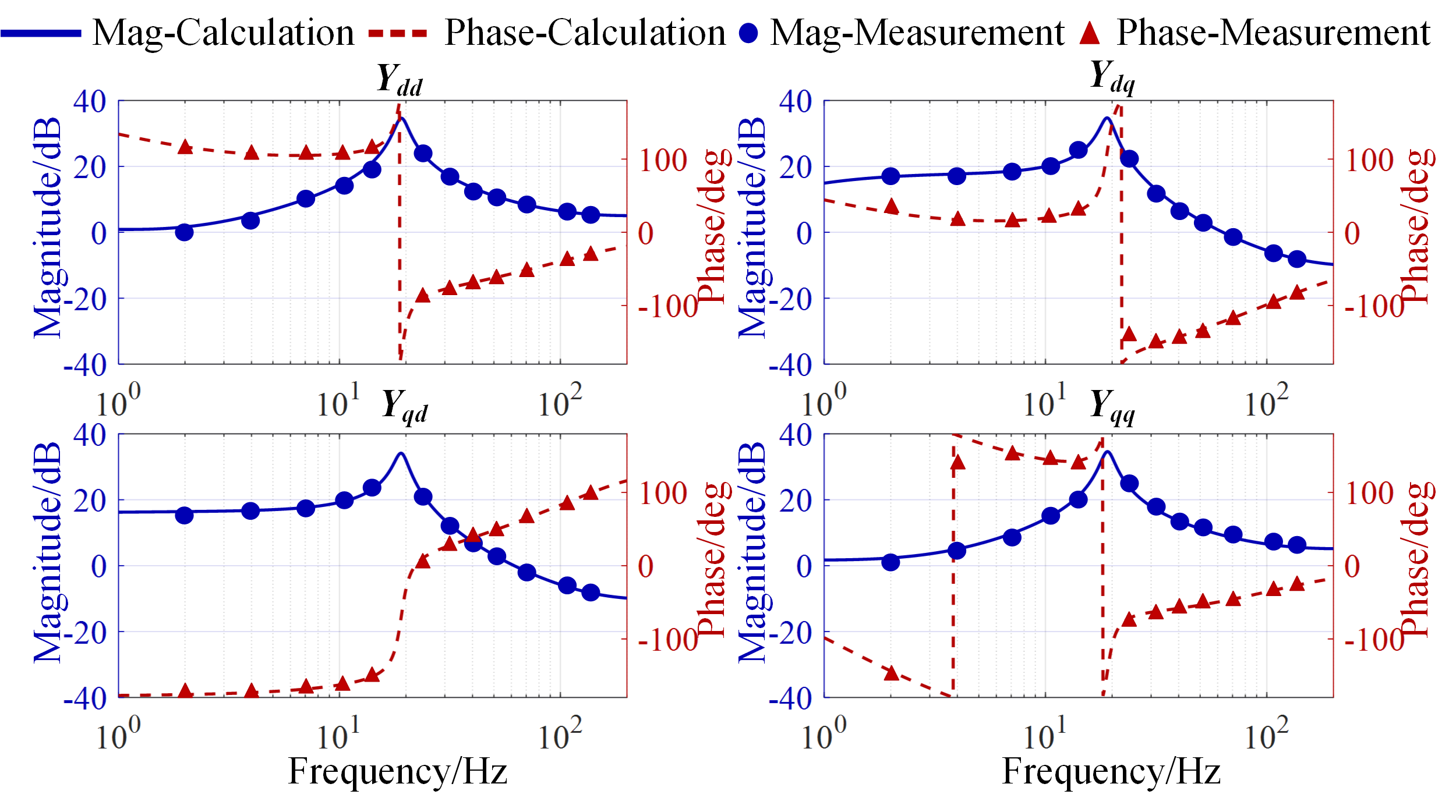}
    \vspace{-15pt}
    \caption{Admittance identification results at rated dispatch commands, with VSC-A at $P_{\mathrm{ref},A} = 100\mathrm{kW}, Q_{\mathrm{ref},A} = 0\mathrm{var}$ as an example.}
    \label{fig:Admittance_Measurement}
    \vspace{-12pt}
\end{figure}


Fig.~\ref{fig:Admittance_Measurement} presents the admittance identification results for VSC-A, which is shown as a representative example, under rated dispatch commands. The proposed method demonstrates high fidelity relative to analytical benchmarks, restricting average magnitude and phase errors to within 1\% and 3$^\circ$, respectively. Minor deviations are confined to the low-frequency band ($<$5~Hz), attributed to inherent measurement errors near the fundamental frequency. 
This precision proves consistent across all heterogeneous units, validating the method's robustness against topological variations.

\subsubsection{Performance of PINN-Based Admittance Extrapolation}

To augment the sparse measurement dataset for comprehensive stability analysis, the PINN model adopts a transfer-learning strategy to extrapolate admittance characteristics over the full operating region. The Measurement Unit (VSC-E), whose parameters are available in the source-domain model, is used to construct the source dataset. A dense source dataset is generated by evaluating the analytical admittance model over a grid of operating points defined by $U_d \in [0.8, 1.2]$~p.u., $U_q \in [-0.2, 0.2]$~p.u., and $I_{d,q} \in [-1, 1]$~p.u.\ with 0.1~p.u.\ resolution. Combined with perturbation frequencies $f_\mathrm{p} \in [1, 150]$~Hz, this yields 2,500 distinct samples.


Hyperparameters are optimized via Bayesian optimization including dropout and learning rates. The pre-trained model achieves a training and validation MSE of ${3.27 \times 10^{-2}}$ and ${6.73\times 10^{-2}}$, respectively, demonstrating effective capture of the underlying physics-based admittance structure.

Subsequently, the model is fine-tuned using the sparse experimental samples (25 operating points $\times$ 12 frequencies) obtained from each target VSC. Centered Kernel Alignment (CKA) analysis~\cite{Korblith2019CKA} reveals that Layers I and II extract universal physical features, exhibiting high similarity scores ($>0.89$) across heterogeneous topologies. Consequently, these layers are frozen to prevent overfitting, while the subsequent layers are updated. This approach enables accurate admittance reconstruction corresponding to the entire feasible dispatch command space ($P_{\mathrm{ref}}, Q_{\mathrm{ref}} \in [-1, 1]$~p.u.), overcoming the limitation of sparse measurements.

\begin{figure}[htbp]
    \centering
    \vspace{-18pt}
    \includegraphics[width=0.9\linewidth]{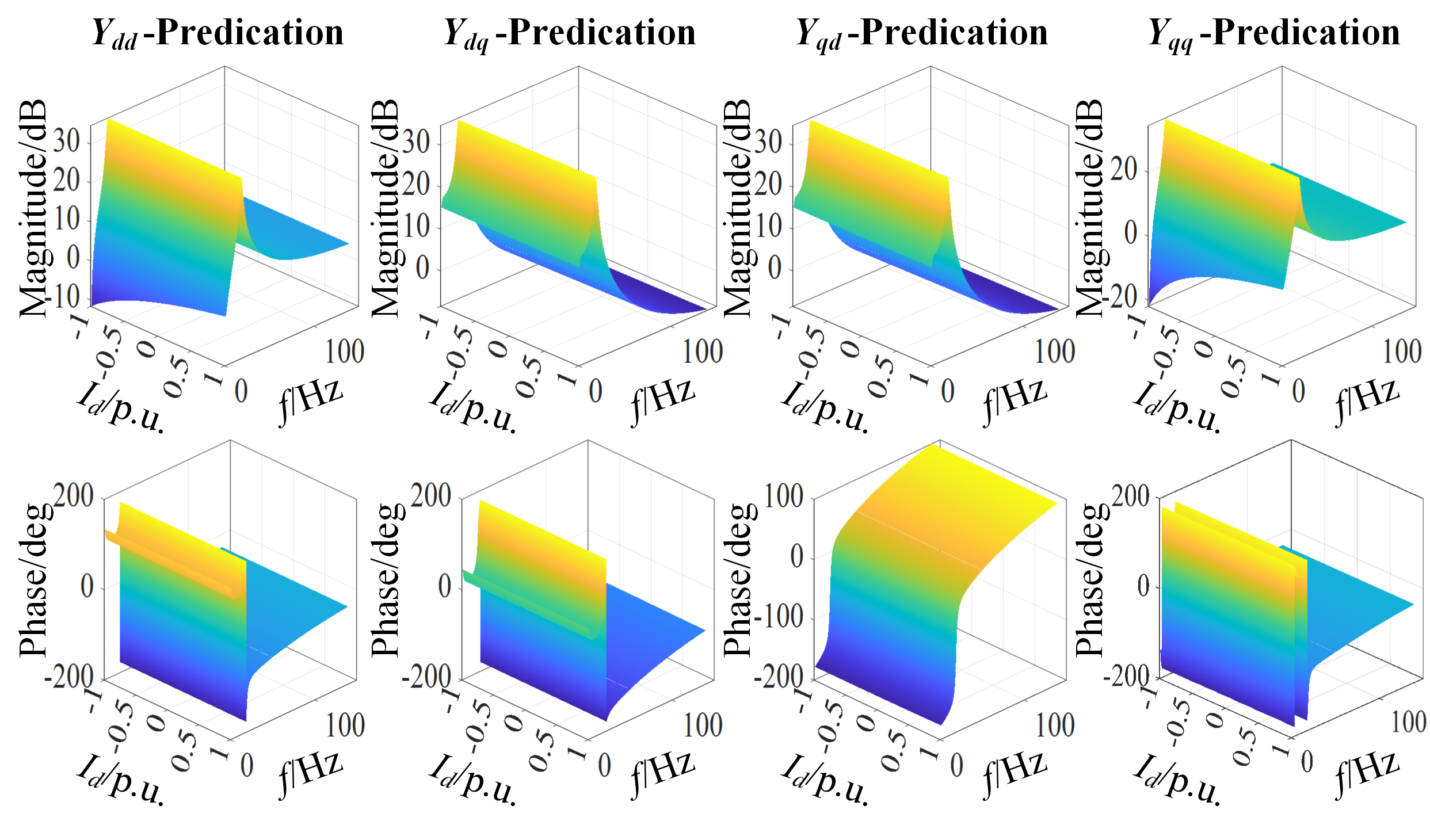}
    \vspace{-15pt}
    \caption{PINN-predicted admittance of VSC-A.}
    \label{fig:VSC-A_Pre}
    \vspace{-1em}
\end{figure}

\begin{figure}[htbp]
    \centering
    \vspace{-2em}
    \includegraphics[width=0.9\linewidth]{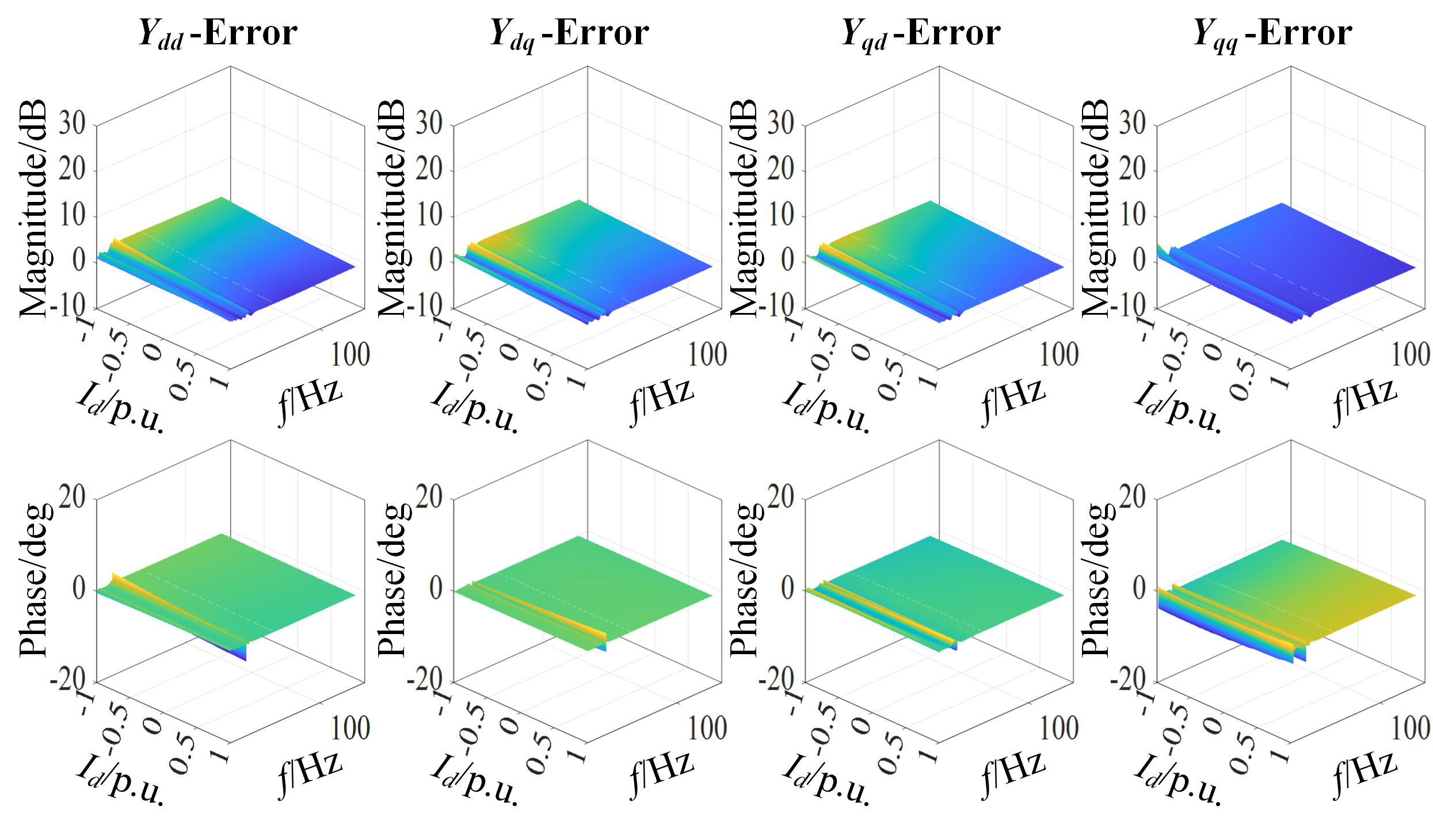}
    \vspace{-15pt}
    \caption{Magnitude and phase prediction errors of PINN model for VSC-A.}
    \label{fig:VSC-A_Error}
    \vspace{-1em}
\end{figure}

Fig.~\ref{fig:VSC-A_Pre} visualizes the PINN-predicted admittance of VSC-A, specifically mapping the response to variations in d-axis current $I_d$ and frequency $f_\mathrm{p}$ while holding other operating states constant. VSC-A is presented here as a representative example for visualization, while the same extrapolation procedure is applied to all target VSCs in the subsequent vulnerability identification. The corresponding error landscape, presented in Fig.~\ref{fig:VSC-A_Error}, corroborates the model's extrapolation fidelity. Prediction errors are consistently suppressed below 3~dB in magnitude and $4^\circ$ in phase across the majority of the operating envelope. Although slightly elevated deviations occur in the low-frequency band, these residuals remain well within the tolerance required for reliable stability margin assessment.

\subsubsection{Identification of the Most Vulnerable Inverter and Worst-Case Dispatch Command}

To implement the stability margin optimization framework in Section~III.D, the uncertainty parameter set $\mathcal{P}$ is first quantified. The virtual impedance parameters $R_{\mathrm{v},i}$ and $L_{\mathrm{v},i}$ are inversely estimated using steady-state terminal measurements, specifically the Nominal Operating Point (NOP) obtained under normal dispatch commands. 
By rearranging the virtual-impedance equations in \eqref{eq:steady_state}, the virtual-impedance parameters are estimated from the nominal operating point as: 
\begin{equation}
\label{eq:virtual_impedance_estimation}
\left\{
\begin{aligned}
R_{v,i} &=
\frac{
I_{d,i}^{\mathrm{NOP}}\!\left(E_{\mathrm{v}}-U_{d,i}^{\mathrm{NOP}}\right)
-
I_{q,i}^{\mathrm{NOP}}U_{q,i}^{\mathrm{NOP}}
}{
\left(I_{d,i}^{\mathrm{NOP}}\right)^2+\left(I_{q,i}^{\mathrm{NOP}}\right)^2
} \\
L_{v,i} &=
\frac{
-
I_{q,i}^{\mathrm{NOP}}\!\left(E_{\mathrm{v}}-U_{d,i}^{\mathrm{NOP}}\right)
-
I_{d,i}^{\mathrm{NOP}}U_{q,i}^{\mathrm{NOP}}
}{
\omega_n\!\left[\left(I_{d,i}^{\mathrm{NOP}}\right)^2+\left(I_{q,i}^{\mathrm{NOP}}\right)^2\right]
}
\end{aligned}
\right.
\end{equation}

This formulation ensures high observability using NOP data. Considering the sensor precision requirements in IEEE Std C37.118.1~\cite{IEEEPMUStd}, a robust uncertainty bound of $\pm 5\%$ is applied to accommodate linearization residuals.
The droop coefficients $D_{q,i}$ are treated as uncertain normalized parameters in the steady-state reactive-power mapping, and are varied within representative bounds consistent with typical Q-V droop design practice and IEEE Std 1547~\cite{IEEE1547Std}.
\vspace{-1.2em}

\begin{figure}[htbp]
    \centering
    \includegraphics[width=0.9\linewidth]{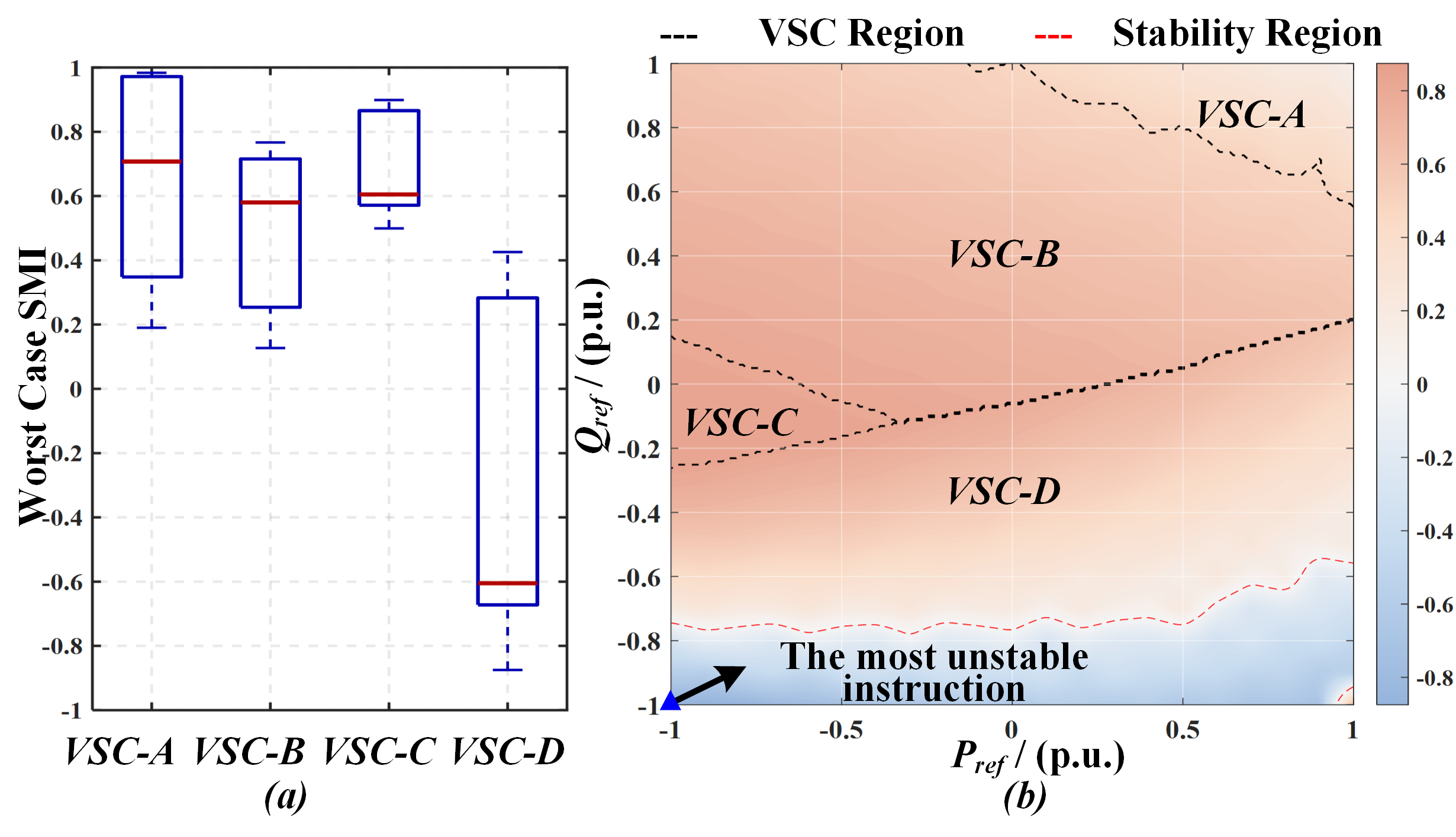}
    \vspace{-16pt}
    \caption{
    SMI-based vulnerability assessment results: (a) Distribution of worst-case SMI for each VSC over the uncertainty set $\mathcal{P}$; (b) Vulnerability region partition in the dispatch-command domain under a representative baseline parameter setting, with $R_v$ and $L_v$ fixed at their NOP-based estimates.
    }
    
    \label{fig:SMI_result}
    \vspace{-0.8em}
\end{figure}

With $\mathcal{P}$ quantified, the worst-case SMI optimization problem~\eqref{eq:optimization} is solved via a grid search over the feasible dispatch command space $\mathcal{O}_k$. The resulting vulnerability distributions, shown in Fig.~\ref{fig:SMI_result}(a), indicate that VSC-D exhibits the widest variability and lowest stability margins across the uncertainty space, indicating the highest vulnerability despite its relatively small power rating $10\mathrm{kVA}$. 

Further detailing the stability landscape, Fig.~\ref{fig:SMI_result}(b) partitions the dispatch command domain $(P_{\mathrm{ref}}, Q_{\mathrm{ref}})$ based on the VSC yielding the minimum SMI, computed under the baseline parameter scenario. The red dashed line delineates the stability boundary (SMI $= 0$), revealing that VSC-D dominates the negative-damping regions.

\begin{table}[htbp]
\centering
\vspace{-1.5em}
\scriptsize
\caption{Identification of the worst-case dispatch commands for stability margin minimization}
\vspace{-1em}
\label{tab:critical_commands}
\begin{tabular}{c @{\hspace{5pt}} cc@{\hspace{5pt}}c}
\hline\hline
Case & Target VSC & Worst SMI & Identified Worst-Case Dispatch Command \\
\hline
1 & None  & 0.902  & None\\
2 & VSC-A & 0.191  & $P_{\text{ref}} = -1.0$ p.u., $Q_{\text{ref}} = 1.0$ p.u.\\
3 & VSC-B & 0.127  & $P_{\text{ref}} = -1.0$ p.u., $Q_{\text{ref}} = 1.0$ p.u.\\
4 & VSC-C & 0.499  & $P_{\text{ref}} = -1.0$ p.u., $Q_{\text{ref}} = 1.0$ p.u.\\
5 & VSC-D & \textbf{-0.875}  & $P_{\text{ref}} = -1.0$ p.u., $Q_{\text{ref}} = -1.0$ p.u.\\
\hline\hline

\end{tabular}
\end{table}

\vspace{-10pt}
As quantitatively summarized in Table~\ref{tab:critical_commands}, the global minimum stability margin (SMI $=-0.875$) is pinpointed at the destabilizing dispatch command for VSC-D ($P_{\mathrm{ref},D}^* = -1.0\mathrm{p.u.}, Q_{\mathrm{ref},D}^*=-1.0\mathrm{p.u.}$). Consequently, this specific command vector constitutes the optimal attack strategy to induce inherent instability.

\vspace{-12pt}
\subsection{Experimental Validation of Dispatch Command Manipulation Attack}

In this section, the effect of the identified worst-case dispatch command attack is validated through CHIL experiments on the platform described in Section IV-A. The experiments focus on three aspects: 1) system-wide instability induced by manipulating the identified most vulnerable VSC, 2) comparison with an existing destabilization method, and 3) validation of identification accuracy across multi-VSC attack scenarios.

\begin{figure}
    \centering
    \includegraphics[width=0.9\linewidth]{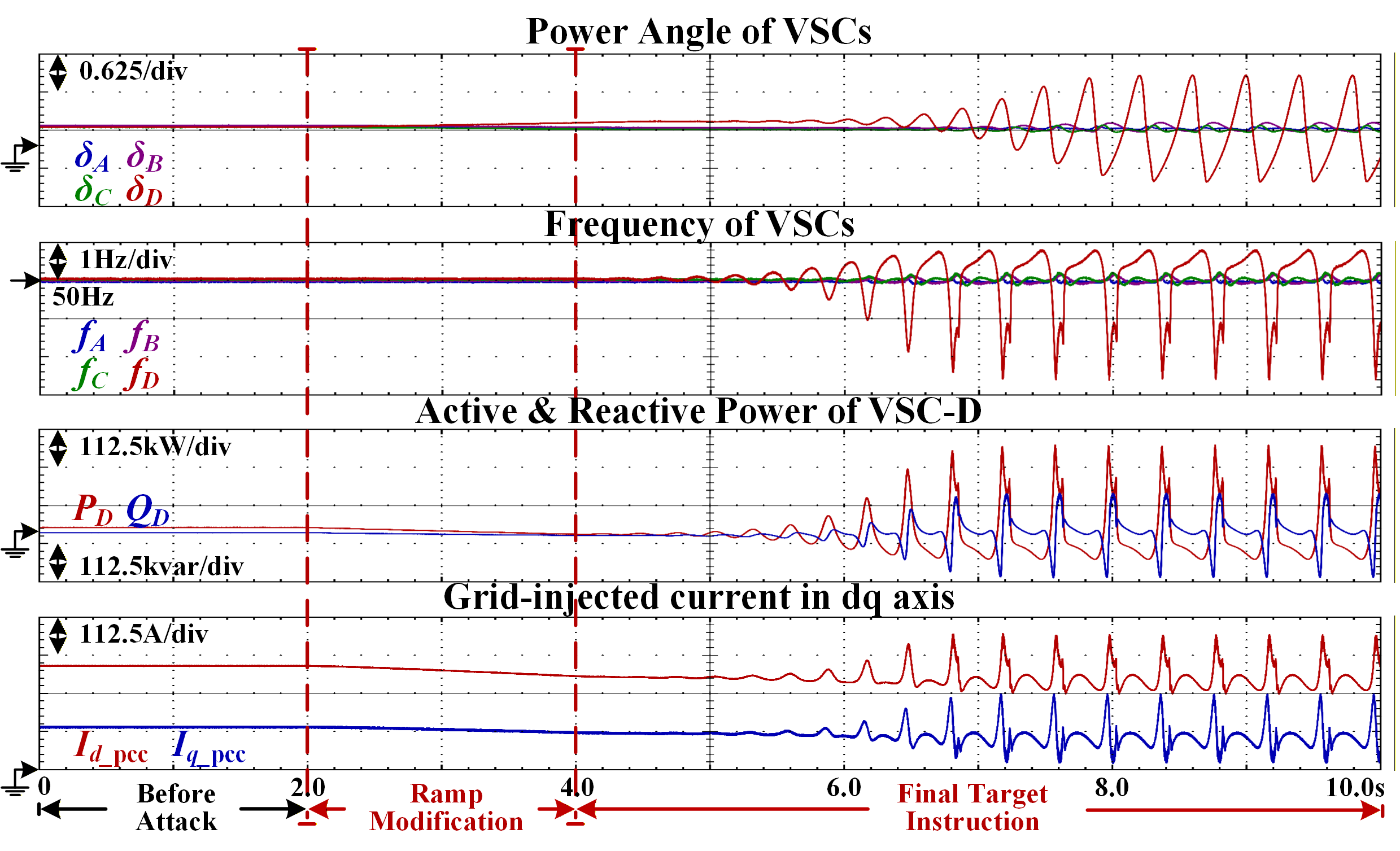}
    \vspace{-1em}
    \caption{Microgrid stability degradation under the proposed targeted attack: (a) power angles of VSCs A--D; (b) frequencies of VSCs A--D; (c) active and reactive power of VSC-D; (d) grid-injected $d$-axis and $q$-axis currents.}
    \label{fig:StraightResult}
    \vspace{-1.2em}
\end{figure}

\begin{figure}
    \centering
    \includegraphics[width=0.9\linewidth]{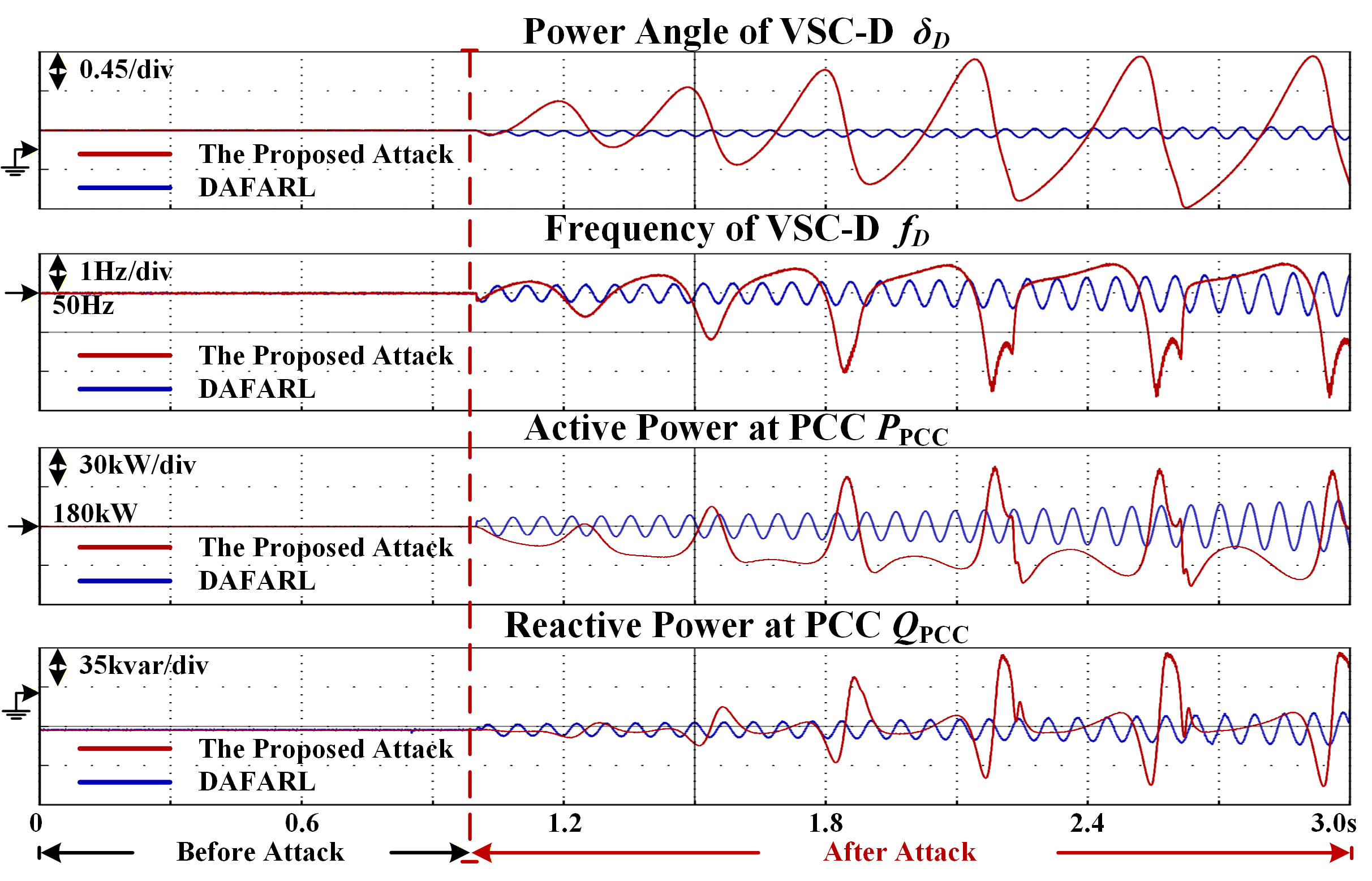}
    \vspace{-1em}
    \caption{Performance comparison between the identified destabilizing dispatch command manipulation and existing methods.}
    \label{fig:comparisons}
    \vspace{-2em}
\end{figure}

\begin{figure*}[b]
    \centering
    \vspace{-2em}
    \includegraphics[width=0.9\linewidth]{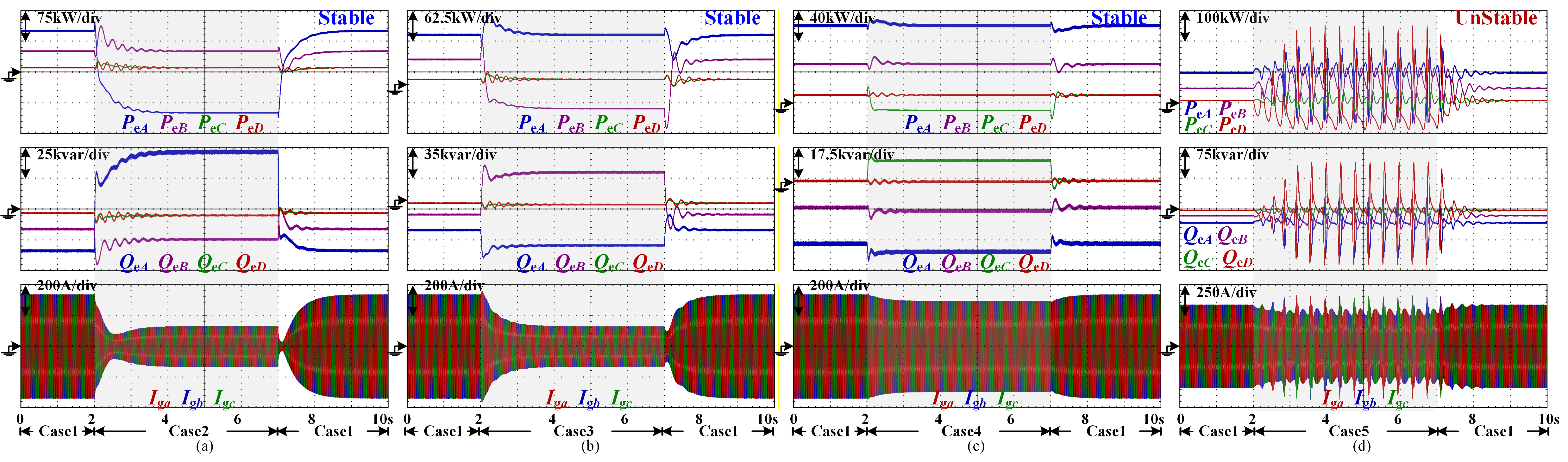}
    \vspace{-15pt}
    \caption{System dynamic responses under worst-case reference manipulation targeting different VSCs: (a)--(c) scenarios where the system remains stable (Cases 2--4); (d) scenario where the system exhibits oscillatory instability (Case 5).}
    \vspace{-20pt}
    \label{fig:PQrefChanged}
\end{figure*}
Fig.~\ref{fig:StraightResult} depicts the system dynamic response under the proposed attack targeting VSC-D, identified as the most vulnerable inverter in Section IV-B. 
At $t = 2~\mathrm{s}$, the reference commands of VSC-D are manipulated to the identified destabilizing dispatch command derived in Table II.
Crucially, a ramp-based modification is employed to rigorously distinguish the identified small-signal instability from transient instability triggered by large disturbances. This gradual transition ensures the system evolves quasi-statically, confirming that the subsequent divergence is driven by the intrinsic negative damping at the target operating point rather than a transient shock.
Consequently, this manipulation precipitates system-wide instability, manifesting as divergent oscillations in power angles, frequencies, active/reactive power outputs, and grid-injected currents. Specifically, the system transitions from stable operation to sustained oscillation within approximately 1.2s, with frequency deviations exceeding ±1 Hz and power angle separations among VSCs growing unboundedly. 

To benchmark the proposed strategy against a representative interaction-driven destabilization method, a comparative analysis is performed against DAFARL~\cite{blackbox_Yu2025DARAFL}, a reinforcement learning based destabilizing attack, as illustrated in Fig.~\ref{fig:comparisons}. To provide a favorable comparison for DAFARL, we assume it has already identified the most sensitive parameter and relax its modification range to [0.2, 1.5] p.u., which exceeds practical feasibility. Even under these relaxed conditions, the proposed strategy produces faster divergence and higher-amplitude oscillations while dispatch command remains entirely within nominal bounds. This result indicates that dispatch-command manipulation constitutes a more practical instability-inducing vector than parameter tampering.

Finally, to further verify the accuracy of the proposed identification procedure, attacks are applied to each VSC individually using the worst-case commands listed in Table~II, and the corresponding dynamic responses are shown in Fig.~\ref{fig:PQrefChanged}. To clearly observe the divergence trajectory, protection mechanisms and limiters are temporarily bypassed. The system operates at the nominal operating point until $t = 2$~s, when the command manipulation is applied, and the commands are restored at $t = 7$~s. 
The results show that only the command targeting VSC-D (Case~5) induces system-wide instability, resulting in 5.42-Hz sub-synchronous oscillations. By contrast, the commands targeting VSC-A/B/C (Cases~2--4), although identified as the worst-case commands for those units, do not compromise the overall system stability. These results confirm that effective destabilization requires accurate identification of the most vulnerable inverter, and that indiscriminate command manipulation on arbitrary VSCs is insufficient even under full control access.

\vspace{-16pt}
\section{Conclusion}

This paper investigates dispatch command manipulation in VSC-dominated smart microgrids and develops an admittance-guided identification framework to locate the most vulnerable inverter and its corresponding worst-case dispatch command. The central observation is that VSCs may exhibit reduced damping margins under certain operating conditions, such that dispatch-command-induced operating-point migration can drive the system toward instability without requiring direct modification of controller parameters. By combining measurement-based admittance identification with physics-informed-neural-network-based extrapolation, the proposed framework reconstructs operating-point-dependent admittance characteristics from terminal measurements and enables stability-margin-oriented vulnerability identification under a limited-prior-knowledge setting. CHIL experiments validate that the identified destabilizing command can induce severe sub-synchronous oscillations, highlighting that effective destabilization depends on precise vulnerability identification rather than indiscriminate manipulation. More broadly, the results reveal the limitations of defense mechanisms based solely on static limit checking and motivate the need for stability-aware command screening in VSC-dominated microgrids.

\vspace{-13pt}
\bibliographystyle{IEEEtran}
\bibliography{mybibfile}

\end{spacing}

\end{document}